\documentclass[
	reprint,
	superscriptaddress,
	showkeys,
	aps,
	pra,
	longbibliography,
	nofootinbib
]{revtex4-2}

\usepackage[utf8]{inputenc}

\usepackage[hyperref]{xcolor}
	\definecolor{goethe-blau}{cmyk}{1.0,0.2,0.0,0.4}
	\definecolor{hellgrau}{cmyk}{0.04,0.04,0.05,0.02}
	\definecolor{sandgrau}{cmyk}{0.12,0.09,0.13,0.0}
	\definecolor{dunkelgrau}{cmyk}{0.25,0.25,0.30,0.75}
	\definecolor{emo-rot}{cmyk}{0.04,1.0,0.8,0.07}
	\definecolor{purple}{cmyk}{0.08,1.0,0.3,0.36}
	\definecolor{senfgelb}{cmyk}{0.01,0.25,1.0,0.05}
	\definecolor{gruen}{cmyk}{0.62,0.4,0.87,0.09}
	\definecolor{magenta}{cmyk}{0.08,0.86,0.12,0.12}
	\definecolor{orange}{cmyk}{0.0,0.7,1.0,0.04}
	\definecolor{sonnengelb}{cmyk}{0.0,0.12,0.95,0.0}
	\definecolor{helles-gruen}{cmyk}{0.4,0.17,0.81,0.07}
	\definecolor{lichtblau}{cmyk}{0.8,0.0,0.06,0.04}

\usepackage[
	colorlinks,
	pdfpagelabels,
	breaklinks,
	pdfstartview=FitH,
	bookmarksopen=true,
	bookmarksnumbered=true,
	bookmarksopenlevel=2,
	plainpages=false,
	hypertexnames=false,
	citecolor=emo-rot,
	linkcolor=goethe-blau,
	urlcolor=purple,
	pdftitle={{Numerical fluid dynamics for FRG flow equations: Zero-dimensional QFTs as numerical test cases. II. Entropy production and irreversibility of RG flows}},
	pdfauthor={{Adrian Koenigstein, Martin J. Steil, Nicolas Wink, Eduardo Grossi and Jens Braun}},
	pdfkeywords={entropy, numerical entropy, total variation diminishing, Functional Renormalization Group, conservation laws, numerical fluid dynamics, irreversibility, Z2 model, zero-dimensional QFT, C-theorem}
]{hyperref}

\usepackage{amsmath} 
\usepackage{amssymb} 
\usepackage{bm} 

\usepackage{bookmark} 
\usepackage{graphicx} 

\usepackage{booktabs} 
\usepackage{dcolumn} 

\allowdisplaybreaks 

\begin{document}


\title{
	Numerical fluid dynamics for FRG flow equations:\texorpdfstring{\\}{ }Zero-dimensional QFTs as numerical test cases.\texorpdfstring{\\}{ }II. Entropy production and irreversibility of RG flows
}

\author{Adrian Koenigstein}
	\email{koenigstein@th.physik.uni-frankfurt.de}
	\affiliation{
		Institut f\"ur Theoretische Physik, Goethe University,\\
		Max-von-Laue-Stra{\ss}e 1, D-60438 Frankfurt am Main, Germany
	}

\author{Martin J.\ Steil}
	\email{msteil@theorie.ikp.physik.tu-darmstadt.de}
	\affiliation{
		Technische Universit\"at Darmstadt, Department of Physics, Institut f\"ur Kernphysik, Theoriezentrum,\\
		Schlossgartenstra{\ss}e 2, D-64289 Darmstadt, Germany
	}

\author{Nicolas Wink}
	\email{wink@thphys.uni-heidelberg.de}
	\affiliation{
		Technische Universit\"at Darmstadt, Department of Physics, Institut f\"ur Kernphysik, Theoriezentrum,\\
		Schlossgartenstra{\ss}e 2, D-64289 Darmstadt, Germany
	}
	\affiliation{
		Institut f\"ur Theoretische Physik, University Heidelberg,\\
		Philosophenweg 16, D-69120 Heidelberg, Germany
	}

\author{Eduardo Grossi}
\email{eduardo.grossi@stonybrook.edu}
\affiliation{
	Center for Nuclear Theory, Department of Physics and Astronomy,\\
	Stony Brook University, Stony Brook, NY 11794, U.S.A.
}

\author{Jens Braun}
	\email{jens.braun@physik.tu-darmstadt.de}
	\affiliation{
		Technische Universit\"at Darmstadt, Department of Physics, Institut f\"ur Kernphysik, Theoriezentrum,\\
		Schlossgartenstra{\ss}e 2, D-64289 Darmstadt, Germany
	}
	\affiliation{
		Helmholtz Research Academy Hesse for FAIR, Campus Darmstadt,\\
		D-64289 Darmstadt, Germany
	}
	\affiliation{
		ExtreMe Matter Institute EMMI, GSI,\\ 
		Planckstraße 1, D-64291 Darmstadt, Germany
	}

\date{\today}

\begin{abstract}
	We demonstrate that the reformulation of renormalization group (RG) flow equations as non-linear heat equations has severe implications on the understanding of RG flows in general. We demonstrate by explicitly constructing an entropy function for a zero-dimensional $\mathbb{Z}_2$-symmetric model that the dissipative character of generic non-linear diffusion equations is also hard-coded in the functional RG equation. This renders RG flows manifestly irreversible, revealing the semi-group property of RG transformations on the level of the flow equation itself.
	Additionally, we argue that the dissipative character of RG flows, its irreversibility and the entropy production during the RG flow may be linked to the existence of a so-called $\mathcal{C}$-/$\mathcal{A}$-function. In total, this introduces an asymmetry in the so-called RG time -- in complete analogy to the thermodynamic arrow of time --  and allows for an interpretation of infrared actions as equilibrium solutions of dissipative RG flows equations. The impossibility of resolving microphysics from macrophysics is evident in this framework.
	
	Furthermore, we directly link the irreversibility and the entropy production in RG flows to an explicit numerical entropy production, which is manifest in diffusive and non-linear partial differential equations (PDEs) and a standard mathematical tool for the analysis of PDEs. Using exactly solvable zero-dimensional $\mathbb{Z}_2$-symmetric models, we explicitly compute the (numerical) entropy production related to the total variation non-increasing property of the PDE during RG flows toward the infrared limit.
	
	Finally, we discuss generalizations of our findings and relations to the $\mathcal{C}$-/$\mathcal{A}$-theorem as well as how our work may help to construct   truncations of RG flow equations in the future, including numerically stable schemes for solving the corresponding PDEs.
\end{abstract}

\keywords{entropy, numerical entropy, total variation diminishing, Functional Renormalization Group, conservation laws, numerical fluid dynamics, irreversibility, $\mathbb{Z}_2$ model, zero-dimensional QFT, $\mathcal{C}$-theorem}

\maketitle

\section{Introduction}
\label{sec:introduction}

Our modern understanding of Quantum Field Theories (QTFs) and particularly phase transitions is built upon the analysis of renormalization group (RG) trajectories.
In fact, RG theory facilitates our understanding by connecting microscopic and macroscopic physics in a continuous manner.
This is often visualized at the example of block spin transformations~\cite{Kadanoff:1966wm,Wilson:1979qg} which provides an intuitive picture of so-called RG flows in position space.
A modern, functional approach to RG theory is provided by the Functional Renormalization Group (FRG).
It allows for non-perturbative studies of QTFs with applications ranging from biophysics over condensed matter to high-energy physics and quantum gravity, see Ref.~\cite{Dupuis:2020fhh} for a recent overview.

	In Refs.~\cite{Grossi:2019urj,Grossi:2021ksl,Koenigstein:2021syz,Steil:2021cbu} it is shown that renormalization group flows can be seen as flows in the literal sense. The RG time $t = - \ln \big( \tfrac{k}{\Lambda} \big)$, where $k$ is the RG scale in units of energy and $\Lambda$ is some ultraviolet (UV) reference scale, can be identified with an abstract time and directions in field space correspond to spatial directions\footnote{A specific example is the RG flow of (the field-space derivative of) a local potential, which can involve advective and diffusive contributions as well as source/sink terms \cite{Grossi:2019urj,Grossi:2021ksl,Wink:2020tnu,WinkHirschegg,Koenigstein:2020Talk,Koenigstein:2021syz,Koenigstein:2021,Steil:2021cbu,Stoll:2021ori,Ihssen2020}.}, \textit{cf.}\ Refs.~\cite{Zumbach:1993zz,Zumbach:1994kc,Zumbach:1994vg,Hasenfratz:1985dm,Felder:1987} for similar identifications in related flow equations. With this at hand, it becomes appealing to look for further connections between the research fields of (numerical) fluid dynamics and RG theory.
	
	Such a connection is discussed in this paper. To be specific, we shall show that the numerical entropy, which is of utmost importance in the theoretical treatment of 
	partial differential equations (PDEs), see, \textit{e.g.}, the textbooks \cite{Lax1973,Ames:1992,LeVeque:1992,LeVeque:2002,Hesthaven2007,Toro2009,RezzollaZanotti:2013}, has a very close connection to an entropy in the RG flow\footnote{In this context we also have to mention the subsequent publication \cite{Cotler:2022fze} by J. Cotler and S. Rezchikov who were able to interpret the Polchinski equation as an ``optimal transport gradient flow of a field-theoretic relative entropy'' thus establishing a firm and explicit connection between an information-theoretic entropy and (F)RG flows.} and further possible connections to the so-called $\mathcal{C}$-/$\mathcal{A}$-functions for RG theories, \textit{cf.}\ Refs.~\cite{Zamolodchikov:1986gt,Rosten:2010vm,Banks:1987qs,Cardy:1988cwa,Osborn:1989td,Jack:1990eb,Komargodski:2011vj,Curtright:2011qg,Haagensen:1993by,Generowicz:1997he,Forte:1998dx,Codello:2013iqa,Codello:2015ana,Becker:2014pea,Becker:2016zcn} and Sec.~\ref{sec:c-theorem_irreversibility_entropy} for more details on $\mathcal{C}$-/$\mathcal{A}$-functions.
	
	One of the most important direct consequences of this is that the same ``(thermodynamic) arrow of time'' or ``thermodynamic time asymmetry'' \cite{Lebowitz:2008} identified by the entropy of a PDE, is also present from an RG perspective.
	
	In nature as well as in the PDEs that describe our physical world, entropy is produced by diffusion (dissipation) as well as discontinuities of all kind. Consequently the evolution of such systems and also their numerical solutions are irreversible and usually only weak solutions are accessible numerically \cite{Lax1973,Ames:1992,LeVeque:1992,LeVeque:2002,Hesthaven2007,Toro2009,RezzollaZanotti:2013}. As we will demonstrate in this paper, the total variation non-increasing property and related numerical entropy, used to guarantee the stability of numeric solution schemes, can be promoted to a ``physical'' entropy function sharing some characteristics with a $\mathcal{C}$-function and its properties transfer from the PDE to the QFT and vice versa. Therefore, RG flows are also not reversible.\footnote{Note that similar arguments, which link the dissipative character of RG flow equations to the irreversibility of the RG flow, were already brought up in Refs.~\cite{Zumbach:1994vg} and \cite{Zamolodchikov:1986gt} already before or parallel to the development of the functional RG framework pioneered in Ref.~\cite{Wetterich:1992yh}.} This makes the semi-group character of the RG, see, \textit{e.g.}, Ref.~\cite{deligne1999quantum}, explicit. This semi-group character also becomes manifest in Kadanoffs block-spin picture \cite{Kadanoff:1966wm,Wilson:1971bg,Wilson:1971dh,Wilson:1979qg}. The irreversibility of RG flows is not just an abstract concept but presents on a practical level in rather simple truncations of the Functional Renormalization Group (FRG) equation. 

	These statements may have no severe practical implications for studies of, \textit{e.g.}, QCD and condensed-matter systems, where the RG flow is in general  followed from small (UV limit) to large length scales (IR limit). In these cases, the dynamics in the long-range limit is predicted from a given known UV action by integrating out high momentum modes along the ``natural'' RG-time direction. However, in situations where RG flows are followed from large to small length scales, such as studies of the asymptotic safety scenario in QFTs (see Refs.~\cite{Weinberg:1976xy,Weinberg:1996kw,Percacci:2007sz,Weinberg:2009ca,Rosten:2010vm}, Refs.~\cite{Niedermaier:2006wt,Bonanno:2020bil} for a recent review in the context of (quantum) gravity, and Refs.~\cite{Braun:2010tt,Jakovac:2014lqa} for applications in condensed-matter physics), the question of irreversibility of RG flows and the associated production of entropy may indeed be very relevant. 
	
	Whereas RG flows are indeed reversible for certain classes of truncations (of the underlying effective action), we shall demonstrate in the present work (with the aid of simple models) that it becomes formally impossible to reverse RG flows in cases where no truncations of the effective action are made. Even more, already for often employed truncation schemes (\textit{e.g.}, local potential approximations), we shall see that irreversibility associated with numerical entropy production can already be a manifest feature of RG flows. Of course, irreversibility of RG flows does not imply that it is not possible to construct theories which are valid on all scales. It only implies that the search for such theories may in general be more complicated. In any case, generalizations of the arguments presented in our present work may help to provide a fresh view on these aspects (and/or revive some already existing arguments \cite{Zamolodchikov:1986gt,Zumbach:1993zz,Zumbach:1994kc,Zumbach:1994vg,Rosten:2010vm,Felder:1987,Hasenfratz:1985dm}).
	
	As we shall discuss below, fixed points still play an important role within the fluid dynamic interpretation of RG flows. In fact, fixed points can be identified with steady-flow solutions and/or (thermal) equilibrium situations on the level of the rescaled dimensionless flow equations, which have advective and diffusive character.

	One major benefit of the connection revealed in this paper is that a measure for the irreversibility of the RG flow is explicitly provided via the identification with numerical entropy and especially total variation \cite{LeVeque:1992,LeVeque:2002,RezzollaZanotti:2013,KTO2-0,HARTEN1983357}. Hence, the construction and analysis of such a measure, at least in certain truncations, might be greatly simplified.\footnote{We note that observations similar to ours have already been pointed out in the works of Refs.~\cite{Rosten:2010vm,Zumbach:1993zz,Zumbach:1994kc,Zumbach:1994vg} for related (partially linearized) flow equations.} In future, this might also help to single out adequate truncation schemes for RG flow equations as those truncations, which maintain the irreversible character of the flow of the full untruncated system.

	This paper is organized as follows: In Sec.~\ref{sec:basics}, we briefly discuss the methodological framework of our present study. This includes both the functional RG approach and its correspondence to fluid dynamics. Moreover, we introduce the zero-dimensional $O(1)$ model which underlies our numerical studies. Numerical entropy and the total variation non-increasing property is then discussed in detail in Sec.~\ref{sec:entropy_c-function_tvd}. Explicit computations and a detailed analysis of numerical entropy production in a variety of test cases are presented in Sec.~\ref{sec:numerical_tests}.
In Sec.~\ref{sec:c-theorem_irreversibility_entropy}, we then give a discussion of the possibility of a generalization of our present findings with respect to the irreversibility of RG flows, entropy production and the \texorpdfstring{$\mathcal{C}$}{C}-theorem to higher-dimensional theories. Finally, our conclusions and a brief outlook can be found in Sec.~\ref{sec:conc}.

\section{Functional RG, fluid dynamics, and the zero-dimensional \texorpdfstring{$O(1)$}{O(1)} model}\label{sec:basics}

\subsection{FRG framework}
	This section is dedicated to a brief summary of the key aspects of the FRG and the zero-dimensional $O(1)$ model within this framework. For a comprehensive discussion, we refer to Part~I of our series of publications on numerical fluid dynamics and FRG flow equation \cite{Koenigstein:2021syz} as well as to Refs.~\cite{Grossi:2019urj,Grossi:2021ksl,Ihssen2020,Wink:2020tnu,Steil:2021cbu} and upcoming publications \cite{Stoll:2021ori,Koenigstein:2021}. For more general reviews on the FRG method, we refer to Refs.~\cite{Pawlowski:2005xe,Rosten:2010vm,Dupuis:2020fhh,Berges:2000ew,Kopietz:2010zz,Gies:2006wv,PawlowskiScript,Delamotte:2007pf}.\\
	
	The FRG framework is built on an exact RG equation \cite{Ellwanger:1993mw,Morris:1993qb,Wetterich:1991be,Wetterich:1992yh}, which is a functional 
	partial integro-differential equation for the scale-dependent effective average action $\bar{\Gamma}_t [ \Phi ]$:
		\begin{align}
			\partial_t \bar{\Gamma}_t [ \Phi ] = \mathrm{STr} \Big[ \big( \tfrac{1}{2} \, \partial_t R_t \big) \, \big( \bar{\Gamma}_t^{(2)} [ \Phi ] + R_t \big)^{-1} \Big] \, .	\label{eq:exact_renormalization_group_equation}
		\end{align}
	The equation holds for arbitrary dimensions and arbitrary field content, which is summarized in the ``super''-field~$\Phi$, \textit{cf.}\ Refs.~\cite{Reuter:1993kw,Reuter:1996cp,Reuter:2001ag,Braun:2007bx}. Here,
		\begin{align}
			t = - \ln \big( \tfrac{k}{\Lambda} \big)	\label{eq:rg_time}
		\end{align}
	is the RG time (note our sign convention), while $k/\Lambda$ constitutes the ratio of the RG scale $k$ and the UV cutoff scale $\Lambda$. The latter is the scale, where the Exact Renormalization Group (ERG) equation is initialized with the classical action ${\bar{\Gamma}_{t = 0} [ \Phi ] = \mathcal{S} [ \Phi ]}$. The ``super''-trace stands for a trace in field space, momentum-/position-space, as well as all internal spaces, \textit{e.g.}, color, flavor \textit{etc.} and $R_t$ denotes an monotonically decreasing scale-dependent IR regulator function, see, \textit{e.g.},  Refs.~\cite{Rosten:2010vm,Litim:2000ci,Pawlowski:2017gxj,Pawlowski:2005xe,Braun:2020bhy,Osborn:2011kw} for details. Solving the ERG equation \eqref{eq:exact_renormalization_group_equation} by integrating the full set of PDEs that can be generated from the ERG via suitable projections, from $t = 0$ to $t \rightarrow \infty$, thus calculating the full quantum IR effective action $\Gamma [ \Phi ] \equiv \bar{\Gamma}_{t \rightarrow \infty} [ \Phi ]$, is equivalent to calculating all $1$PI-$n$-point-correlation (vertex) functions via a partition function/functional integral \cite{Weinberg:1996kr,Peskin:1995ev,ZinnJustin:2002ru,Iliopoulos:1974ur,DeWitt:1965jb,Greiner:1996zu,Wipf:2013vp,PawlowskiScript,Pawlowski:2005xe}. The ERG equation \eqref{eq:exact_renormalization_group_equation} is the direct mathematical implementation of Wilson's idea of the Renormalization Group \cite{Wilson:1971bg,Wilson:1971dh,Polchinski:1983gv}: Obtaining the macrophysics from the micro-physics via gradually integrating out momentum-shells from the UV to the IR, which corresponds to a coarse-graining process in position space, \textit{e.g.}, Kadanoff's block-spin transformations \cite{Wilson:1979qg,Kadanoff:1966wm,Delamotte:2007pf}. Earlier formulations of the RG in terms of similar flow equations can be found in Refs.~\cite{Wegner:1972ih,Hasenfratz:1985dm,Polchinski:1983gv,Brydges:1987,Zumbach:1993zz,Zumbach:1994kc,Zumbach:1994vg}.

\subsection{The zero-dimensional \texorpdfstring{$O(1)$}{O(1)}-model}

	For what concerns this paper, we study one of the probably most simplistic QFTs imaginable within this framework -- a zero-dimensional $\mathbb{Z}_2$ symmetric model or $O(1)$ model\footnote{Although being technically imprecise, we mainly refer to the model as the $O(1)$ model, thus the special case $N = 1$ of an $O(N)$ symmetry, where the symmetry is only realized in terms of discrete transformations.}. Still, as the interested reader will experience by studying this and the parallel publications of our series \cite{Koenigstein:2021syz,Steil:2021cbu}, as well as Refs.~\cite{Hikami:1978ya,Bessis:1980ss,DiVecchia:1990ce,Nishigaki:1990sk,Schelstraete:1994sc,Zinn-Justin:1998hwu,Fl_rchinger_2010,Moroz:2011thesis,Keitel:2011pn,Strocchi:2013awa,Kemler:2013yka,Pawlowski:talk,Rentrop_2015,Rosa:2016czs,Liang:2017whg,SkinnerScript,Millington:2019nkw,Alexander:2019cgw,Catalano:2019,Millington:2020Talk,Millington:2021ftp,Kades:2021hir}, this model (and its extension, the zero-dimensional $O(N)$ (vector) model) is non-trivial and can serve as a minimalistic tool to highlight and test fundamental features of and basic methods for QFTs, not only for pedagogical purposes, but also as real benchmark scenarios.
	
	For the zero-dimensional $O(1)$ model, the most general ``truncation'' to solve the ERG equation \eqref{eq:exact_renormalization_group_equation} is the local potential approximation
		\begin{align}
			\bar{\Gamma}_t [ \varphi ] = U ( t, \varphi ) \, ,
		\end{align}
	where $U ( t, \varphi )$ is simply a function of $t$ and the real scalar (mean) field $\varphi$ that is called the \textit{scale-dependent effective potential}. Space-time or momentum-space dependences of the field or integrations over the former as well as derivatives of the field do not exist. The entire QFT consists of a scalar $\phi$ (field), which can assume arbitrary real values and can be thought of as a self-interacting ``particle'' in a single point. The theory is therefore maximally coupled and ultra-local. The only additional requirement, which we impose on ${\langle\phi\rangle=\varphi}$ and $U ( t, \varphi )$, is that the (mean) field transforms as follows
		\begin{align}
			\varphi \mapsto \varphi^\prime = - \varphi
		\end{align}
	under $\mathbb{Z}_2$-transformations and that $U ( t, \varphi )$ is in turn invariant under these transformations
		\begin{align}
			U ( t, \varphi ) = U ( t, - \varphi ) \, .
		\end{align}
	The RG flow from $t = 0$ to $t \rightarrow \infty$ is initialized with the classical action (here the classical potential),
		\begin{align}
			\bar{\Gamma}_{t = 0} ( \varphi ) = U ( t = 0 , \varphi ) = \mathcal{S} ( \varphi ) = U ( \varphi ) \, ,
		\end{align}
	which also represents an ordinary function of $\varphi$ with $\mathbb{Z}_2$-symmetry. Still, in order to render the corresponding partition function as well as the expectation values \eqref{eq:expectation_values} convergent, $U ( \varphi )$ has to be bounded from below and, for $|\varphi|\rightarrow\infty$, it has to grow at least with $\varphi^2$. Nevertheless, there is no need to claim smoothness or analyticity for $U ( \varphi )$, as discussed in detail in part I and III of this series of publications~\cite{Koenigstein:2021syz,Steil:2021cbu} and Sec.~\ref{sec:numerical_tests} of this work.
	
	The corresponding ERG equation \eqref{eq:exact_renormalization_group_equation} simplifies drastically for the zero-dimensional $O(1)$ model,\footnote{Note that we evaluated $\varphi$ on a ``background-field configuration'' $\sigma$, which is actually not needed for the zero-dimensional $O(1)$ model because the field $\varphi$ is already space-time-independent. Still, we adopt the conventions used in Ref.~\cite{Koenigstein:2021syz} and higher-dimensional scenarios.}
			\begin{align}
				\partial_t U ( t, \sigma ) = \frac{\tfrac{1}{2} \, \partial_t r ( t )}{r ( t ) + \partial_\sigma^2 U ( t, \sigma )} = %
				\begin{gathered}
					\includegraphics{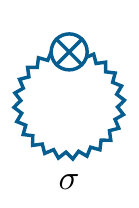}
				\end{gathered} \, .	\label{eq:o(1)_wetterich_equation}
			\end{align}
	However, it retains its fundamental one-loop structure and all its characteristic properties as a non-linear parabolic PDE and initial value problem in one ``temporal'' $t \in [0, \infty)$ and one ``spatial'' $\sigma \in ( - \infty, \infty )$ direction. Furthermore, it remains an exact equation without any truncation.
	
	A major difference to higher-dimensional $O(1)$ models (\textit{e.g.}, when using the LPA-optimized regulator \cite{Litim:2000ci,Pawlowski:2017gxj}) is the absence of an additional $t$-dependent factor on the right hand side of the equations, \textit{cf.}\ Refs.~\cite{Wetterich:1991be,Wetterich:1992yh,Schaefer:2001cn}, which, however, does not conceptually spoil any of our further reasoning. For what follows, we always use a zero-dimensional version of the monotonically decreasing exponential regulator
			\begin{align}
				r ( t ) = \Lambda \, \mathrm{e}^{- t} \,.	\label{eq:exponential_regulator}
			\end{align}
	Here, the UV cutoff must be chosen sufficiently large, \textit{cf.}\ Refs.~\cite{Koenigstein:2021syz,Braun:2018svj,Cichutek:2020bli}. A peculiar feature of the zero-dimensional version of the ERG equation \eqref{eq:exact_renormalization_group_equation} is that an integration to $t \rightarrow \infty$ is indeed possible (which can be seen via reparametrization of the RG time and there is no need for a numerical IR cutoff \cite{Keitel:2011pn,Koenigstein:2021syz}). Nevertheless, we will use non-vanishing IR cutoffs for our numerical calculations in Sec.~\ref{sec:numerical_tests}, to be as close as possible to higher dimensional scenarios.
	
	Having performed the $t$-integration down to the IR limit, 
	we can extract the vertex functions $\Gamma^{(2n)}$ at the physical point\footnote{There is no (spontaneous) symmetry breaking in zero dimensions \cite{Moroz:2011thesis,Koenigstein:2021syz}. This is a consequence of a special version of the Coleman-Mermin-Wagner-Hohenberg theorem \cite{Mermin:1966,Hohenberg:1967,Coleman:1973ci} or on the level of Eq.~\eqref{eq:expectation_values} simply a consequence of the discrete $\mathbb{Z}_2$-symmetry.} $\langle\phi\rangle=\varphi = \sigma = 0$ by taking (numerical) derivatives of $U ( t = 0, \sigma )$ \textit{w.r.t.}\ $\sigma$. These vertex functions are in direct relation to the expectation values, see Refs.~\cite{Keitel:2011pn,Koenigstein:2021syz},
		\begin{align}
			&	\langle \phi^{2 n} \rangle = \frac{\int_{-\infty}^\infty \mathrm{d} \phi \, \phi^{2n} \, \mathrm{e}^{- \mathcal{S} ( \phi )}}{\int_{-\infty}^\infty \mathrm{d} \phi \, \mathrm{e}^{- \mathcal{S} ( \phi )}} \, ,	&&	n \in \mathbb{N}_0 \, ,	\label{eq:expectation_values}
		\end{align}
	which can be calculated numerically up to machine precision (or sometimes even be evaluated analytically). (Here, $\phi$ denotes the ``fluctuating quantum field''.) This fact makes zero-dimensional QFTs an interesting test ground because numerical FRG calculations can be compared against easily attainable exact results from (numerical) integration of Eq.~\eqref{eq:expectation_values}, \textit{cf.}\ Refs.~\cite{Hikami:1978ya,Bessis:1980ss,DiVecchia:1990ce,Nishigaki:1990sk,Schelstraete:1994sc,Zinn-Justin:1998hwu,Fl_rchinger_2010,Moroz:2011thesis,Keitel:2011pn,Strocchi:2013awa,Kemler:2013yka,Pawlowski:talk,Rentrop_2015,Rosa:2016czs,Liang:2017whg,SkinnerScript,Millington:2019nkw,Alexander:2019cgw,Catalano:2019,Millington:2020Talk,Millington:2021ftp,Kades:2021hir}.
	
\subsection{Fluid-dynamic reformulation of the RG flows}

	In this section, we briefly summarize the main findings of our parallel and upcoming publications \cite{Koenigstein:2021syz,Steil:2021cbu,Koenigstein:2021,Stoll:2021ori} and Refs.~\cite{Grossi:2019urj,Grossi:2021ksl} on the reformulation of the RG flow equation in terms of a fluid-dynamical conservation law.
	
	Taking a derivative \textit{w.r.t.}\ $\sigma$ of the flow equation \eqref{eq:o(1)_wetterich_equation}, we obtain a scalar one-dimensional (here parabolic) conservation law,
				\begin{align}
					\partial_t u ( t, x ) = \frac{\mathrm{d}}{\mathrm{d} x} \, \bigg( \big[ \tfrac{1}{2} \, \partial_t r ( t ) \big] \, \frac{1}{r ( t ) + \partial_x u ( t , x )} \bigg) \, ,	\label{eq:flow_equation_conservative}
				\end{align}
	where $\sigma = x$ is identified with a spatial dimension and $u ( t, x )\equiv \partial_x U(t,x)$ is the conserved quantity. In fact, Eq.~\eqref{eq:flow_equation_conservative} is a non-linear diffusion/heat equation,\footnote{In part I of this series of publications \cite{Koenigstein:2021syz} we demonstrate that the conservative formulation generalizes to zero-dimensional $O(N)$ models, which turn out to be a non-linear advection-diffusion equations, and can even be generalized to higher dimensional models involving fermions in terms of advection-diffusion-source/sink equations \cite{Grossi:2019urj,WinkHirschegg,Wink:2020tnu,Ihssen2020,Koenigstein:2020Talk,Grossi:2021ksl,Koenigstein:2021syz,Steil:2021cbu,Koenigstein:2021,Stoll:2021ori}.}${}^{,}$\footnote{Note that the similarities between RG flow equations and the heat equation were already observed before, \textit{cf.}\ Refs.~\cite{Brydges:1987,Zumbach:1994kc,Rosten:2010vm,SkinnerScript,Salmhofer:2020Talk}, but did -- to the best of our knowledge -- never result in a comprehensive picture.} which can actually be generalized to arbitrary dimension. This conservative formulation and interpretation in terms of (numerical) fluid dynamics has tremendous consequences and benefits for understanding and solving the RG flow equation:
		\begin{enumerate}
			\item	Conservative formulations of RG flow equations provide direct access to the highly developed toolbox of numerical fluid dynamics.
			
			\item	An interpretation of the RG flow equations as flow equations in the narrow sense of the word makes the dynamics during the flow intuitively understandable. Advective contributions (pions in the $O(N)$-scenario) transport the conserved quantity $u ( t, x )$ along the field space direction $\sigma = x$ (bulk motion) and can cause non-analyticities like shocks and rarefaction waves in field space \cite{Grossi:2019urj,Grossi:2021ksl,Aoki:2017rjl}, as is well known for non-linear hyperbolic conservation laws \cite{Rankine:1870,Hugoniot:1887,Lax1973,Ames:1992,LeVeque:1992,LeVeque:2002,Hesthaven2007,Toro2009,RezzollaZanotti:2013}. The non-linear diffusive contribution (the radial sigma mode) smears out cusps and jumps in $u ( t, x )$ and corresponds to undirected movement of $u ( t, x )$ depending on the local ``concentration differences'', the gradient $\partial_x u ( t, x )$, via a highly non-linear diffusion coefficient. On the level of the LPA approximation, also fermionic contributions to the flow can be easily understood in this framework in terms of source/sink terms, see ,\textit{e.g.} Ref.~\cite{Stoll:2021ori}.
		\end{enumerate}
	The reformulation of the flow equation \eqref{eq:o(1)_wetterich_equation} as a diffusion equation \eqref{eq:flow_equation_conservative} has direct implications for the goal of the present paper. As outlined in our parallel discussion in Sec.~IV of Ref.~\cite{Koenigstein:2021syz}, diffusion is one specific dissipative process. Dissipative processes go hand in hand with entropy production and irreversibility.\footnote{This argument also generalizes to the $O(N)$ model involving advection and the large-$N$ limit \cite{Tetradis:1995br}. The corresponding flow equation in the limit $N\rightarrow\infty$ is a purely hyperbolic advection equation. Interacting and arising non-analyticities like shocks and rarefaction waves in non-linear advection equations are sources of entropy and as such lead so irreversible flows, \textit{cf.}\ Refs.~\cite{Grossi:2019urj,Grossi:2021ksl,Steil:2021cbu,Aoki:2014,Aoki:2017rjl} for examples of non-analytical dynamics in RG flows.} We can therefore conclude that the irreversibility of the RG transformations during the RG flow is hard coded in the diffusive character of the ERG equation \eqref{eq:exact_renormalization_group_equation}, not only in zero space-time dimensions, but for any dimension and any QFT, \textit{cf.}\ Refs.~\cite{Zumbach:1994kc,Zumbach:1994kc,Zumbach:1994vg,Zumbach:1993zz}. Hence, the rise of entropy during the RG flow might therefore be directly linked to $\mathcal{C}$-/$\mathcal{A}$-theorems. This is explained in the next sections in the context of our minimalistic toy model QFT.

\section{(Numerical) entropy and the total variation}
\label{sec:entropy_c-function_tvd}

	The first part of this section deals with the explicit construction of a (numerical) entropy for the conservation law \eqref{eq:flow_equation_conservative}. This entropy has to be a functional of the conserved quantity $u ( t, x )$ and/or its derivatives\footnote{The purely diffusive character of Eq.~\eqref{eq:flow_equation_conservative} is expected to smoothen $u$ during the RG flow which renders $u ( t, x )$ differentiable (but not necessarily analytic) at least for $0 < t < \infty$. This does not need to be the case for hyperbolic conservation laws where taking derivatives of $u ( t, x )$ has to be handled with great care, \textit{e.g.}, around shocks.} that is monotonically rising during the RG flow. Monotonicity is explicitly proven for valid initial conditions $U ( t = 0, x )$. Since $u ( t, x )$ is by definition a function of all couplings of the theory, the (numerical) entropy function might therefore be linked to a zero-dimensional version of $\mathcal{C}$-/$\mathcal{A}$-function. In fact it might have some practical advantages compared to some practical approaches toward $\mathcal{C}$-/$\mathcal{A}$-functions studied in literature, since $u ( t, x )$ does not even need to be expandable in explicit couplings at all, but still contains all degrees of freedom.
	
	In the second part of this section, we derive a discrete formulation of this entropy functional and demonstrate that it can be directly related to the \textit{total variation} (TV) of $u ( t, x )$ and the total variation diminishing/non-increasing property (TVD/TVNI) of commonly used numeric schemes for conservation laws \cite{HARTEN1983357,Lax1973,KTO2-0}, which is, why we denote it as a ``numerical'' entropy.

\subsection{Construction of the (numerical) entropy}
\label{subsec:entropy}

	The construction of our (numerical) entropy function is directly inspired by the construction of entropy/energy functionals for the Bateman-Burgers equation \cite{Bateman1915,Burgers1948} or the heat-equation \cite{Cannon:1984}.

	Let $y \in \mathbb{R}$ and
		\begin{align}
			&	s : \mathbb{R} \to \mathbb{R} \, ,	&	y \mapsto s ( y ) \, ,
		\end{align}
	be a continuously twice differentiable convex function on $\mathbb{R}$, hence
		\begin{align}
			&	s ( y ) \in C^2 ( \mathbb{R} ) \, ,	&	s^{\prime\prime} ( y ) \geq 0 \, ,	\label{eq:convexity_of_entropy_function}
		\end{align}
		for all~$y  \in \mathbb{R}$. 
	Furthermore, we require that $s ( y )$ shall not grow faster than $y^2$ for $|y| \rightarrow \infty$, which is explained below. Using $s ( y )$ we define the functional
		\begin{align}
			S [ f ( x ) ] \equiv - \int_{- \infty}^{\infty} \mathrm{d} x \, s ( f ( x ) ) \, ,	\label{eq:entropy_functional}
		\end{align}
	to which we shall refer as \textit{entropy functional}. In general, the bounds of integration are chosen according to the domain of our problem at hand. Next, we prove that, choosing $f ( x ) = \partial_x u ( t, x )$, Eq.~\eqref{eq:entropy_functional} indeed plays the role of an (numerical) entropy for 
	the partial differential equation~\eqref{eq:flow_equation_conservative}. Hence, it measures similarly to $\mathcal{C}$-/$\mathcal{A}$-functions for the RG flows the degrees of freedom and irreversibility. To this end, we explicitly demonstrate that $S [ \partial_x u ( t, x ) ]$ is a monotonically increasing during the RG flow, thus being a monotonic function on $t \in [ 0, \infty )$:
		\begin{align}
			\tfrac{\mathrm{d}}{\mathrm{d} t} \, S [ \partial_x u ( t, x ) ] \geq 0 \, .	\label{eq:entropy_rises}
		\end{align}
	The only further ingredient, which is needed for the proof is the spatial derivative of the flow-equation \eqref{eq:flow_equation_conservative}:
		\begin{align}
			& \partial_t [ \partial_x u ( t, x ) ] =	\vphantom{\bigg(\bigg)}
			\\
			= & - \frac{\mathrm{d}}{\mathrm{d} x} \, \bigg( \big[ \tfrac{1}{2} \, \partial_t r ( t ) \big] \, \frac{\partial_x^2 u ( t, x )}{[ r ( t ) + \partial_x u ( t , x ) ]^2} \bigg) \, .	\nonumber
		\end{align}
	Taking spatial derivatives of $u ( t, \sigma )$ should be allowed at any $t \in ( 0 , \infty )$ because of the smoothening character of the diffusion -- at least in zero space-time dimensions.\footnote{The generalization of this argument to higher-dimensional $O(N)$-type models might be delicate, because the non-linear diffusion can also cause non-analyticities in potentials in the IR if these end up in the symmetry broken phase. Here it might be unavoidable to base and repeat the entire discussion using a rigorous weak/integral formulation of the PDEs under consideration.} Only for $t = 0$ the initial condition may violate smoothness, see part I of this series of publications \cite{Koenigstein:2021syz} for a detailed discussion of this subtle issue. Below, we normalize the entropy and subtract the entropy of the initial condition $S [ \partial_x u ( t = 0, x ) ]$, such that this should not spoil any of our subsequent arguments.
	
	Let us now evaluate Eq.~\eqref{eq:entropy_rises}:
\begin{widetext}
		\begin{align}
			& \tfrac{\mathrm{d}}{\mathrm{d} t} \, S [ \partial_x u ( t, x ) ] =	\vphantom{\Bigg(\Bigg)}
			\\
			= \, & - \tfrac{\mathrm{d}}{\mathrm{d} t} \, \int_{-\infty}^{\infty} \mathrm{d} x \, s ( \partial_x u ( t, x ) ) =	\vphantom{\Bigg(\Bigg)}	\nonumber
			\\
			= \, & - \int_{-\infty}^{\infty} \mathrm{d} x \, \big( \partial_t [ \partial_x u ( t, x ) ] \big) \, s^\prime ( \partial_x u ( t, x ) ) =	\vphantom{\Bigg(\Bigg)}	\nonumber
			\\
			= \, & \int_{-\infty}^{\infty} \mathrm{d} x \, \bigg[ \frac{\mathrm{d}}{\mathrm{d} x} \, \bigg( \big[ \tfrac{1}{2} \, \partial_t r ( t ) \big] \, \frac{\partial_x^2 u ( t, x )}{[ r ( t ) + \partial_x u ( t , x ) ]^2} \bigg) \bigg] \, s^\prime ( \partial_x u ( t, x ) ) =	\vphantom{\Bigg(\Bigg)}	\nonumber
			\\
			= \, & \int_{-\infty}^{\infty} \mathrm{d} x \, \big[ - \tfrac{1}{2} \, \partial_t r ( t ) \big] \, \frac{[\partial_x^2 u ( t, x )]^2}{[ r ( t ) + \partial_x u ( t , x ) ]^2} \, s^{\prime\prime} ( \partial_x u ( t, x ) ) + \bigg[ \big[ \tfrac{1}{2} \, \partial_t r ( t ) \big] \, \frac{\partial_x^2 u ( t, x )}{[r ( t ) + \partial_x u ( t , x )]^2} \, s^{\prime} ( \partial_x u ( t, x ) ) \bigg]_{-\infty}^{\infty}	\vphantom{\Bigg(\Bigg)}	\nonumber
		\end{align}
\end{widetext}
	Next, we analyze both terms in the last line separately.
		\begin{enumerate}
			\item	We note that all factors in the integrand of the first term are greater or equal to zero: For the regulator insertion, we have 
				\begin{align}
					- \tfrac{1}{2} \, \partial_t r ( t ) \geq 0 \, ,
				\end{align}
			because $r ( t )$ is a monotonically decreasing function. The numerator and the denominator are obviously positive. In fact, for the denominator of the fraction
				\begin{align}
					r ( t ) > \partial_x u ( t, x ) \, ,
				\end{align}
			for all $t$ anyhow, as long as the initial condition $u(t=0,x)$ and the UV cutoff $\Lambda$ are chosen accordingly, \textit{cf.} Ref.~\cite{Koenigstein:2021syz}. Finally,
				\begin{align}
					s^{\prime\prime} ( \partial_x u ( t, x ) ) \geq 0 \, ,
				\end{align}
			holds by construction according to Eq.~\eqref{eq:convexity_of_entropy_function}.
			
			In total, we find that the integrand of the first term is always greater or equal to zero, which directly transfers to the integral itself.
			
			\item	For the second term, we first use that, for large $| x |$, the potential $U ( t, x )$ and all its derivatives do not change during the RG flow, see also part I of this series of publications \cite{Koenigstein:2021syz}. Furthermore, we use that $s ( y )$ maximally grows like $y^2$ for $|y| \rightarrow \infty$ by definition. 
			This implies that its derivative $s^\prime ( y )$ increases asymptotically as $y$ at most. Additionally, we use that $U ( t, x )$ is at least proportional to $x^2$ for $|x| \rightarrow \infty$ in order to have well-defined expectation values \eqref{eq:expectation_values}. Consequently, we have to distinguish two scenarios. If $U ( t, x ) \sim x^2$ for large $|x|$, the second term vanishes identically, due to the third spatial derivative of $U ( t, x )$, namely $\partial_x^2 u ( t, x )$, in the numerator. Otherwise, if $U ( t, x )$ grows faster than $x^2$ for large $|x|$, the denominator $[ r ( t ) + \partial_x u ( t, x ) ]^2$ will always grow faster than the product $[\partial_x^2 u ( t, x )] \, s^\prime ( \partial_x u ( t, x ) )$ for $|x| \rightarrow \infty$. We conclude that the second term always vanishes, provided that the initial conditions come with the assumed large-$|x|$-asymptotic behavior.
		\end{enumerate}
	In total, we have shown the statement of Eq.~\eqref{eq:entropy_rises}, which promotes $S$ to an entropy (functional) of our system 
	that can only increase.\\

	For what follows, we choose the twice differentiable convex function $s ( y ) = y^2$. This implies
		\begin{align}
			S [ \partial_x u ( t, x ) ] = - \int_{- \infty}^{\infty} \mathrm{d} x \, [ \partial_x u ( t, x ) ]^2 \, .	\label{eq:entropy_func}
		\end{align}
	$S$ can be viewed as measure for the richness of structure of the potential -- the information encoded in the potential -- by integrating the square of the gradient of $u(t,x)$ over all positions $x$ in field space.
	
	With the definition \eqref{eq:entropy_func} the following practical problem arises: For practical purposes $S [ \partial_x u ( t, x ) ]$ formally diverges at any time $t$ because $\partial_x u ( t, x )$ is at least constant for $|x| \rightarrow \infty$. This problem can be cured, by subtracting the entropy $S [ \partial_x u ( t = 0, x ) ]$ of the initial condition. Since $u ( t, x )$ does not change for large $|x|$ during the entire RG flow, the infinite but constant contributions cancel and we can observe the relative rise in entropy. This should be a valid approach, since we are only interested in these relative changes anyhow. We therefore define and consider the normalized entropy
		\begin{align}
			\mathcal{C} [ \partial_x u ( t, x ) ] = S [ \partial_x u ( t, x ) ] - S [ \partial_x u ( t = 0, x ) ] \, , \label{eq:Cdef}
		\end{align}
	which is finite. The alphabetic character ``$\mathcal{C}$'' is chosen because this function quantifies irreversibility similarly to $\mathcal{C}$-/$\mathcal{A}$-functions. We are aware of the fact that a real $\mathcal{C}$-/$\mathcal{A}$-function should be based on the dimensionless rescaled flow equation. This issue is discussed in Sub.Sec.~\ref{subsec:ctheorem}.
	
	Eq. \eqref{eq:Cdef} for the $\mathcal{C}$-function makes the loss of information/richness of structure of the effective potential $u(t,x)$ during RG time evolution explicit. $\mathcal{C}$ monotonically increases with RG time $t$ because the richness of structure/information decreases with $t$. A loss of information about a system (effective potential $u(t,x)$) during RG time evolution goes hand in hand with the impossibility to reconstruct/recover earlier states of the system (which had more information) and thus the RG time evolution is irreversible. In the present setup the purely diffusive flow equation of the zero-dimensional $O(1)$ model is responsible for this loss of information as large gradients are smeared out by diffusion during RG time evolution, \textit{cf.} Sec. IV for explicit numerical examples.
	
\subsection{Discrete formulation and relation to the total variation non-increasing property}
\label{subsec:discrete_c_and_tvd}

	In this subsection we discuss a discretized version of Eq.~\eqref{eq:Cdef}, which is suited for practical computations. In the following and without loss of generality we consider a finite volume (FV) discretization \cite{LeVeque:1992,LeVeque:2002,RezzollaZanotti:2013,KTO2-0} of $u ( t, x )$ in $x$ with $n$ volume cells of constant width $\Delta x$, centered at $x_i$, $i \in \{ 0, 1, \ldots, n - 1\}$, see our parallel discussion in Sec.~IV of Ref.~\cite{Koenigstein:2021syz} for details. The RG flow is described by the temporal evolution of the $n$ volume averages $\bar{u}_i ( t )$, which are formally defined as the spatial averages of $u ( t, x )$ over $[ x_{i - \frac{1}{2}}, x_{i + \frac{1}{2}} ]$, where $x_{i \pm \frac{1}{2}} \equiv x_i \pm \tfrac{\Delta x}{2}$.
	For the purpose of calculating $\mathcal{C} [ \partial_x u ( t, x ) ]$, we reconstruct the first derivatives from the set of volume averages $\{\bar{u}_i(t)\}$ by a first order finite difference (FD) forward stencil,
		\begin{align}
			\partial_x u ( t, x_i ) = \frac{\bar{u}_{i + 1} ( t ) - \bar{u}_{i} ( t )}{\Delta x} + \mathcal{O} (\Delta x) \, .	\label{eq:centeredDifferences}
		\end{align}
	For the scope of this work this has proven sufficient since the purely diffusive character of the PDE smoothens $u ( t, x )$.\footnote{However, for non-smooth/non-differentiable initial conditions at $t = 0$, such as Eqs.~\eqref{eq:testing_scenario_non-analytic_quadaratic_asymptote} and \eqref{eq:testing_scenario_4} of our numeric examples, a naive finite difference stencil is of course generically ill-conditioned at the discontinuities. As a direct consequence, the absolute value of $S [ \partial_x u ( t = 0, x ) ]$ strongly depends on the explicit discretization points and the ``capturing of the discontinuity'' in the respective volume cells. For $t \rightarrow \infty$ (as a direct consequence of the Coleman-Mermin-Wagner-Hohenberg theorem \cite{Mermin:1966,Hohenberg:1967,Coleman:1973ci}), $u ( t, x )$ is smooth and the finite difference approximation is well-behaved as long as $\Delta x$ is not too small. We conclude that the absolute values of our entropy function \eqref{eq:Cdef} will strongly depend on $\Delta x$ for non-differentiable initial conditions in the IR because we use $S [ \partial_x u ( t = 0, x ) ]$ as normalization, while the qualitative behavior (monotonic rise) is independent of the discretization, which is also true for the discrete total variation \eqref{eq:TVdiscrete}. For the smooth initial conditions \eqref{eq:testing_scenario_phi4} and \eqref{eq:testing_scenario_phi6}, we observed little dependence of the absolute values of the $\mathcal{C}[\partial_x u ( t, x )]$ on $\Delta x$, as expected. Similar discussions will arise for $O(N)$-type models in higher-spacetime dimensions, when their RG flows end in the symmetry broken phase with a non-analytic IR-potential.} 
	
	We use a grid with the first volume cell of the computational domain centered at zero, $x_0 = 0$, and the last centered at a finite $x_\mathrm{max}$, hence $x_{n - 1} = x_\mathrm{max}$. $x_\mathrm{max}$ is chosen large enough, such that $u ( t, x_\mathrm{max} ) = u ( t = 0, x_\mathrm{max} )$ holds to a sufficient level for all $t$, compare our discussion in Ref.~\cite{Koenigstein:2021syz} as well as Refs.~\cite{Grossi:2019urj,Pangon:2009pj,Borchardt:2015rxa,Caillol:2012zz}. This enables a computation of $\mathcal{C} [ \partial_x u ( t, x ) ]$ considering only $x \in [ - x_\mathrm{max}, + x_\mathrm{max} ]$ since the difference $S [ \partial_x u ( t, x ) ] - S [ \partial_x u ( t = 0, x ) ]$ practically vanishes for $|x|\geq x_\mathrm{max}$. 	We therefore study the following quantity:
		\begin{align}
			\mathcal{C} [ \partial_x u ( t, x ) ] = \, & - 2 \int_{0}^{x_\mathrm{max}} \mathrm{d} x \, \big[ \partial_x u ( t, x ) \big]^2 +	\vphantom{\bigg(\bigg)}
			\\
			& + 2 \int_{0}^{x_\mathrm{max}} \mathrm{d} x \, \big[ \partial_x u ( t = 0, x ) \big]^2	\vphantom{\bigg(\bigg)},	\nonumber
		\end{align}
	leveraging the $\mathbb{Z}_2$-symmetry of the problem at hand.
	Inserting Eq.~\eqref{eq:centeredDifferences} and performing the integrals over the constant segments in the volume cells leads to our semi-discrete formulation
		\begin{align}
			\mathcal{C} [ \{\bar{u}_i(t)\} ] = \, & - \frac{2}{\Delta x} \Bigg( \sum_{i = 0}^{n-1} \frac{[ \bar{u}_{i + 1} ( t ) - \bar{u}_{i} ( t ) ]^2}{( 1 + \delta_{i,0} + \delta_{i,n-1} ) } - \label{eq:Cdiscrete}
			\\
			& - \sum_{i = 0}^{n-1} \frac{ [ \bar{u}_{i + 1} ( 0 ) - \bar{u}_{i} ( 0 ) ]^2}{( 1 + \delta_{i,0} + \delta_{i,n-1} )} \Bigg) \, ,	\nonumber
		\end{align}
	where the factor $(1+\delta_{i,0}+\delta_{i,n-1})$ takes into account the fact that we only integrate over the right half of the first and the left half of the last volume cell. 
	
	Practical computations of solutions to the PDE \eqref{eq:flow_equation_conservative} on the compact interval $x \in [ 0, x_\mathrm{max} ]$ require carefully chosen boundary conditions \cite{Koenigstein:2021syz,Steil:2021cbu} to be consistent with solutions of the pure initial value problem posed by Eq.~\eqref{eq:flow_equation_conservative} on the interval $x \in ( - \infty, + \infty )$ \cite{Borchardt:2015rxa,Borchardt:2016pif}. In the present finite volume setup we implement boundary conditions with ``ghost cells" at $x_{-2}$, $x_{-1}$, $x_{n}$ and $x_{n+1}$, where the corresponding cell averages are chosen due to the $\mathbb{Z}_2$-anti-symmetry of $u ( t, x )$ in cases of $\bar{u}_{-2} ( t )$ and $\bar{u}_{-1} ( t )$ and by means of linear extrapolation in the cases of $\bar{u}_{n} ( t )$ and $\bar{u}_{n+1} ( t )$, see Sub.Sec.~IV D of part I of this series of publications \cite{Koenigstein:2021syz} for details. For the computation of $\mathcal{C} [ \{ \bar{u}_i ( t ) \} ]$ we require only the ghost-cell average $\bar{u}_{n} ( t ) = 2 \bar{u}_{n-1} ( t ) - \bar{u}_{n-2} ( t )$ as well as the cell averages at $x_i$ for $i \in \{ 0, 1, \ldots, n - 1 \}$. \\
	
	The entropy functional \eqref{eq:entropy_func} introduced in Sub.Sec.~\ref{subsec:entropy} is closely related to the total variation \cite{HARTEN1983357} -- which is simply the arc length -- of the solution $u ( t, x )$,
		\begin{align}
			\mathrm{TV} [ \partial_x u ( t, x ) ] \equiv \int_0^{x_\mathrm{max}} \mathrm{d} x \, | \partial_x u ( t, x ) | \, ,	\label{eq:TVcontinuous}
		\end{align}
	on the (computational) interval $[ 0, x_\mathrm{max} ]$. The TV qualitatively differs only by a global sign from the entropy functional $S$, where the sign used for the TV is compatible with the mathematical convention for (numerical) entropy. The use of the absolute value $| \partial_x u ( t, x ) |$ in Eq.~\eqref{eq:TVcontinuous} instead of the square $[ \partial_x u ( t, x ) ]^2$ used for $S$ presents only as quantitative difference, which is not of any practical relevance in this work. 
	
	On a FV grid, a typical discretized version of Eq.~\eqref{eq:TVcontinuous} is given by, \textit{cf.}\ Refs.~\cite{HARTEN1983357,LeVeque:1992,LeVeque:2002,RezzollaZanotti:2013},
		\begin{align}
			\mathrm{TV} [ \{ \bar{u}_i ( t ) \} ] \equiv \sum_{i = 0}^{n-1} | \bar{u}_{i+1} ( t ) - \bar{u}_{i} ( t ) | \, ,	\label{eq:TVdiscrete}
		\end{align}
	where a first order forward FD stencil is used to discretize the first derivative. The differences $\mathcal{C} [ \{ \bar{u}_i ( t^{m} ) \} ] - \mathcal{C} [ \{ \bar{u}_i ( t^{m + 1} ) \} ]$ and $\mathrm{TV} [ \{ \bar{u}_i ( t^{m + 1} ) \} ] - \mathrm{TV} [ \{ \bar{u}_i ( t^{m} ) \} ]$ on a discrete trajectory $\bar{u}_i ( t )$ of an admissible solution at different times separated by one time step $\Delta t$, where $t^{m+1} = t^m + \Delta t$, are both greater or equal to zero for all $t^m$. Thus total variation -- arc length -- is non-increasing\footnote{In literature total variation diminishing (TVD) is often used as a less precise synonym for total variation non-increasing (TVNI), \textit{cf.} Sec.~9.2.2 of Ref.\ \cite{RezzollaZanotti:2013}.} and the corresponding entropy in the sign convention of this paper is non-decreasing -- monotonically increasing.\\
	
	(Weak) solutions of broad classes of hyperbolic and parabolic conservation laws are total variation non-increasing during time evolution when considered on a finite interval, see, \textit{e.g.}, Refs.~\cite{HARTEN1983357,LeVeque:1992,Toro2009} and especially Ref.~\cite{Redheffer1974Mar}. The flow Eq.~\eqref{eq:flow_equation_conservative} under consideration in this paper is a non-linear, parabolic pure diffusion equation and the construction of the normalized entropy functional $\mathcal{C}$ of Eq.~\eqref{eq:Cdef} can be adapted to prove directly that solutions of the flow Eq.~\eqref{eq:flow_equation_conservative} are TVNI. The notion of numerical entropy (and TV as a possible candidate for it) is very important in the study, construction and numerical computation of physical weak solutions of conservative equations, see, \textit{e.g.}, the textbooks \cite{Lax1973,Ames:1992,LeVeque:1992,LeVeque:2002,Hesthaven2007,Toro2009,RezzollaZanotti:2013} for further details.

\section{Numerical entropy production in zero-dimensional models}
\label{sec:numerical_tests}

	In this section we present explicit numerical results for the RG flows of the (numerical) entropy function \eqref{eq:Cdef} for some selected zero-dimensional $O(1)$ models (different UV initial conditions). As examples, we choose the test cases which are introduced and discussed in great detail in Sec.~V of part I of this series of publications \cite{Koenigstein:2021syz}. All information on the explicit numerical treatment is presented in Sec.~IV of part I of this series of publications \cite{Koenigstein:2021syz}, where the Kurganov-Tadmor (KT) central scheme \cite{KTO2-0} is discussed and applied to RG flow equations. The numerical parameters for the RG flows of this paper are stated in the figures and their respective captions. An elaborated discussion on the choice and tests of numerical and model parameters can also be found in Chap.~V of part I of this series of publications \cite{Koenigstein:2021syz}. For the sake of completeness and as proof of reliability of our numerical scheme and the choice of our numerical parameters, we nevertheless provide a comparison in Tab.~\ref{tab:sc_1_n_point_functions_exact} between numerical results for the 1PI-two-point-function $\Gamma^{(2)}$ calculated via the solution of the flow equation \eqref{eq:flow_equation_conservative} with the KT-scheme and ``exact'' results calculated via expectation values \eqref{eq:expectation_values} from the partition function.
	
	Note that all plots of the entropy in this section are based on a direct implementation of Eq.~\eqref{eq:Cdiscrete}.
	
		\begin{table}[b]
			\caption{\label{tab:sc_1_n_point_functions_exact}%
				The table lists the ``exact'' results for $\Gamma^{(2)}$ of the $O(1)$ model (second column) for the various UV initial potentials of our test cases (first column), which are calculated by a brute force high-precision one-dimensional numerical integration of the expectation values \eqref{eq:expectation_values} using \textit{NIntegrate} in \texttt{Mathematica} \cite{Mathematica:12.1} with a \textit{PrecisionGoal} and \textit{AccuracyGoal} of $10$. Here, we shall present the first ten digits. The last column lists the relative errors of the numerical solution of the RG flow equation with diffusion obtained with the second order accurate KT-central scheme \cite{KTO2-0} using the parameters listed in the corresponding Figs.~\ref{fig:sc_i_on=1_n=800_xmax=10_lambda=1.0e6_tir=60_rg_flowsc_i_on=1_n=800_xmax=10_lambda=1.0e6_tir=60_rg_flow}, \ref{fig:sc_ii_n_on=1_n=800_xmax=10_lambda=1.0e12_tir=60_rg_flow}, \ref{fig:sc_ii_p_on=1_n=800_xmax=10_lambda=1.0e12_tir=60_rg_flow}, \ref{fig:sc_iii_on=1_n=800_xmax=10_lambda=1.0e12_tir=60_rg_flow} and \ref{fig:sc_iv_on=1_n=800_xmax=10_lambda=1.0e8_tir=60_rg_flow}, see also Ref.~\cite{Koenigstein:2021syz} for a detailed discussion of such errors.
			}
			\begin{ruledtabular}
				\setlength\extrarowheight{2pt}
				\begin{tabular}{l c c}
					UV potential													&	$\Gamma^{(2)}$		&	$| \Gamma^{(2)}_\mathrm{FRG}/\Gamma^{(2)} - 1 |$		\\
					\colrule
					Eq.~\eqref{eq:testing_scenario_non-analytic_quadaratic_asymptote}	&	$0.1768130358$		&	$6.0 \cdot 10^{-6}$\\
					Eq.~\eqref{eq:testing_scenario_phi4} (neg.~mass)					&	$0.1995098930$		&	$1.1 \cdot 10^{-5}$		\\
					Eq.~\eqref{eq:testing_scenario_phi4} (pos.~mass)					&	$1.3324252475$		&	$1.4 \cdot 10^{-5}$		\\
					Eq.~\eqref{eq:testing_scenario_phi6}								&	$0.1740508127$		&	$2.5 \cdot 10^{-5}$		\\
					Eq.~\eqref{eq:testing_scenario_4}									&	$0.2046977422$		&	$5.8 \cdot 10^{-6}$
				\end{tabular}
			\end{ruledtabular}
		\end{table}

\subsection{Test case I: Non-analytic initial condition}
\label{subsec:scenario_I}

\begin{figure}
			\centering
			\includegraphics{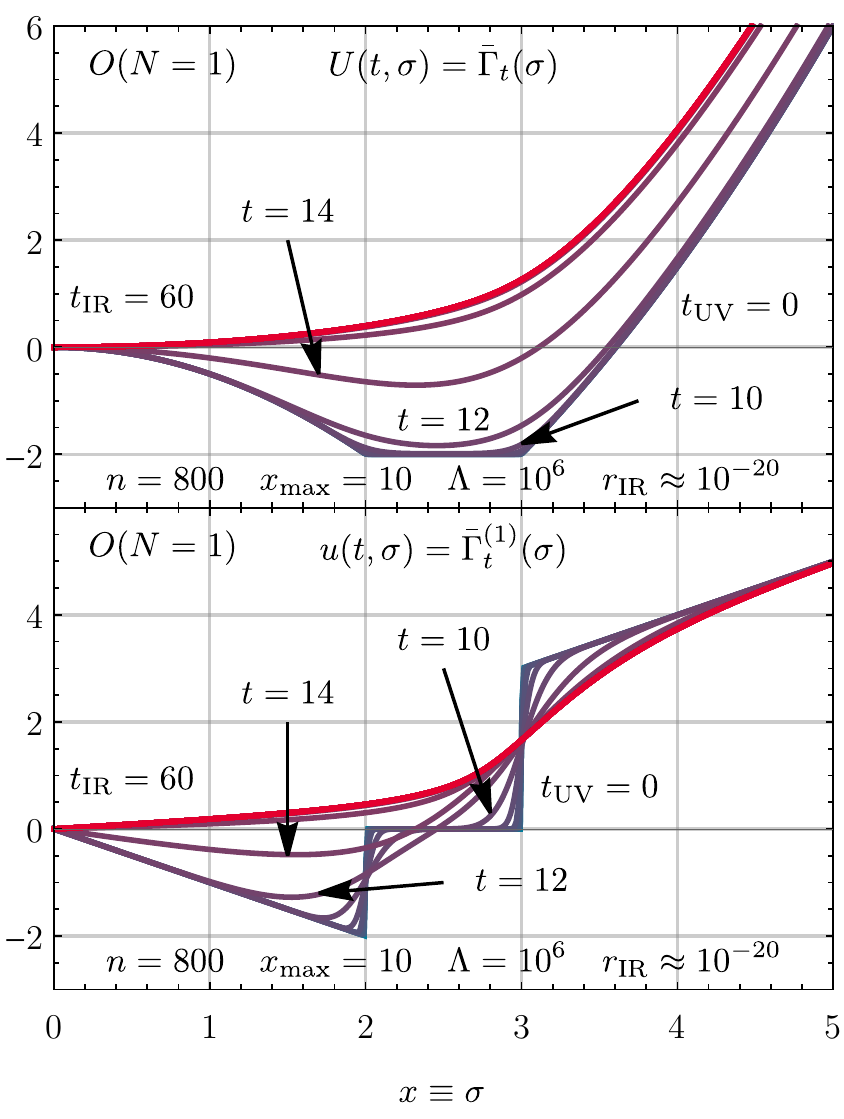}
			\caption{\label{fig:sc_i_on=1_n=800_xmax=10_lambda=1.0e6_tir=60_rg_flowsc_i_on=1_n=800_xmax=10_lambda=1.0e6_tir=60_rg_flow}%
				RG flow of the effective potential $U ( t , \sigma )$ (upper panel) and its derivative $u ( t , \sigma ) = \partial_\sigma U ( t , \sigma )$ (lower panel) for the zero-dimensional $O ( N = 1 )$-model with initial condition Eq.~\eqref{eq:testing_scenario_non-analytic_quadaratic_asymptote} evaluated at $t = 0, \, 2, \, 4, \, \ldots, \, 60$ (integer values for $t$ were only chosen for convenience and readability). The {blue} curve corresponds to the UV while the {red} curve to the IR. We used the exponential regulator Eq.~\eqref{eq:exponential_regulator} with UV cutoff $\Lambda = 10^6$. For convenience only, the plot does not show the region $x = 5$ to $x = 10$ because the tiny differences between $u ( t , \sigma )$ and $u ( t_\mathrm{UV} , \sigma )$ are not visible in this region and vanish for large $x = \sigma$ anyhow. The lower panel is identical to Fig.~8 (upper panel) of Ref.~\cite{Koenigstein:2021syz}.
			}
		\end{figure}

	As our first test case, we choose a UV potential associated with a broken $\mathbb{Z}_2$ symmetry. Moreover, it shall come with  
	infinitely many degenerate minima as well as non-analytic points at $|\sigma| = 2$ and $|\sigma| = 3$,
		\begin{align}
			U ( \sigma ) =
			\begin{cases}
				- \tfrac{1}{2} \, \sigma^2 \, ,			&	\text{if} \quad |\sigma| \leq 2 \, ,	\vphantom{\bigg(\bigg)}
				\\
				- 2 \, ,									&	\text{if} \quad 2 < |\sigma| \leq 3 \, ,	\vphantom{\bigg(\bigg)}
				\\
				+ \tfrac{1}{2} \, ( \sigma^2 - 13 ) \, ,	&	\text{if} \quad 3 < |\sigma| \, ,	\vphantom{\bigg(\bigg)}
			\end{cases}	\label{eq:testing_scenario_non-analytic_quadaratic_asymptote}
		\end{align}
	see Fig.~4 of Ref.~\cite{Koenigstein:2021syz} and also top panel of Fig.~\ref{fig:sc_i_on=1_n=800_xmax=10_lambda=1.0e6_tir=60_rg_flowsc_i_on=1_n=800_xmax=10_lambda=1.0e6_tir=60_rg_flow} 
	for visualizations. This UV potential amounts to a piecewise linear discontinuous initial condition $u ( t = 0, x )$ for the RG flow equations \eqref{eq:flow_equation_conservative}. The corresponding RG flow of $u ( t, x )$ is presented in Fig.~\ref{fig:sc_i_on=1_n=800_xmax=10_lambda=1.0e6_tir=60_rg_flowsc_i_on=1_n=800_xmax=10_lambda=1.0e6_tir=60_rg_flow}.
	The diffusive character of the $\sigma$-mode is clearly visible from the fact that it smoothens the discontinuities at $x = 2$ and $x = 3$, without any directed propagation (advection) of the conserved quantity $u ( t, x )$. As discussed in Refs.~\cite{Koenigstein:2021syz,Moroz:2011thesis}, the system has to restore the  $\mathbb{Z}_2$ symmetry in the ground state as dictated the Coleman-Mermin-Wagner-Hohenberg theorem \cite{Mermin:1966,Hohenberg:1967,Coleman:1973ci}. In particular, the potential has to become convex \cite{Wipf:2013vp,Fujimoto:1982tc}. This can be directly observed in the plot of the RG flow and read off from Tab.~\ref{tab:sc_1_n_point_functions_exact} -- the two-point function is positive at $\sigma = 0$.
	
\begin{figure}
			\centering
			\includegraphics{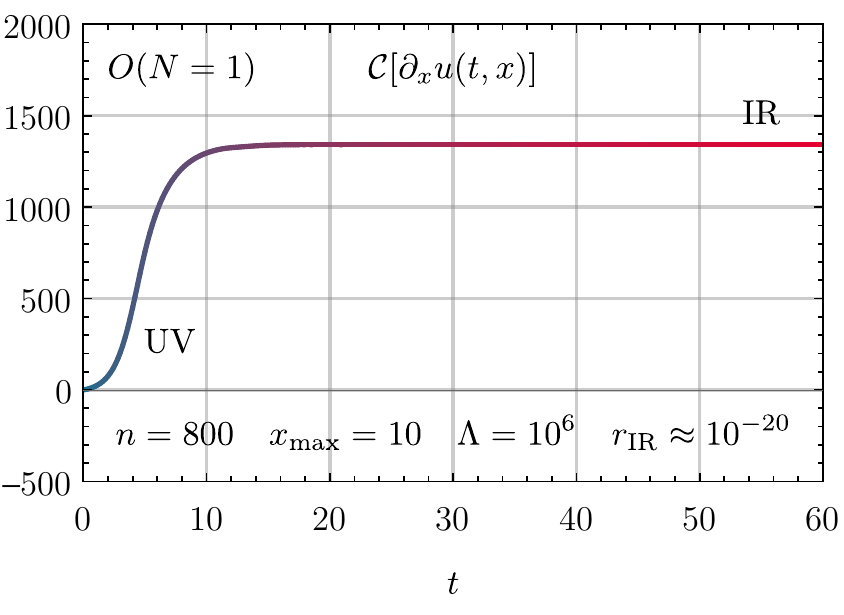}
			\caption{\label{fig:sc_i_on=1_n=800_xmax=10_lambda=1.0e6_tir=60_entropy_flow}%
				The plot shows the monotonic growth of the (numerical) entropy/the $\mathcal{C}$-function $\mathcal{C} [ \partial_x u ( t, x ) ]$ during the RG flow of the test case \eqref{eq:testing_scenario_non-analytic_quadaratic_asymptote} and corresponds to Fig.~\ref{fig:sc_i_on=1_n=800_xmax=10_lambda=1.0e6_tir=60_rg_flowsc_i_on=1_n=800_xmax=10_lambda=1.0e6_tir=60_rg_flow}.
			}
		\end{figure}

	In Fig.~\ref{fig:sc_i_on=1_n=800_xmax=10_lambda=1.0e6_tir=60_entropy_flow} we present the RG flow of the (discretized numerical) entropy function for our first test case.
	As expected from our discussion in Sec.~\ref{sec:entropy_c-function_tvd}, the entropy grows monotonically. It increases by two orders of magnitude starting at zero in the UV until it reaches (again) a plateau in the IR. We find that the entropy grows most when the regulator \eqref{eq:exponential_regulator} reaches the model scales. Loosely speaking, this is where most of the dynamics takes place, see Fig.~\ref{fig:sc_i_on=1_n=800_xmax=10_lambda=1.0e6_tir=60_rg_flowsc_i_on=1_n=800_xmax=10_lambda=1.0e6_tir=60_rg_flow} (approximately between $t \approx 4$ and $t \approx 16$). This is the RG time frame in which the diffusion smears out the discontinuities. From a fluid and thermodynamic perspective and directly on the level of the PDE, the whole process is intuitively understandable: Diffusion goes hand in hand with strong dissipation and a loss of information about the initial state of the system -- the UV, \textit{cf.} Ref.~\cite{Zamolodchikov:1986gt,Zumbach:1994vg}. This is directly comparable to heat conduction, where the information about the initial temperature distribution gets lost during the flow toward ``thermal'' equilibrium \cite{Cannon:1984,LeVeque:1992,Lebowitz:2008}. In the RG framework, this translates to integrating out degrees of freedom from the UV to the IR and a growth in the number of coupling constants in $U ( t, \sigma )$, which is directly related to the growth of entropy. The entropy plateau in the IR is identified with the interacting IR regime and an ``thermal'' equilibrium on the level of the diffusive PDE, whereas a plateau in the UV is associated with a Gaussian UV fixed point \cite{Zinn-Justin:2010,ZinnJustin:2002ru}. As expected the entropy stops changing at these points. IR solutions therefore correspond either to steady-flow solutions (in advection dominated systems for a large number of ``Goldstone'' modes \cite{Nambu:1960tm,Goldstone:1961eq,Goldstone:1962es}) or to (thermal) equilibrium solutions (in diffusion dominated $O(1)$-symmetric systems) in the fluid dynamical picture \cite{Koenigstein:2021syz}.
	
	Note that $t \in [ 0, 60 ]$ corresponds to a integration over $26$ orders of magnitude in $r(t)$, starting $6$ orders of magnitude above the model scales (which are of order one) and ending up $20$ orders of magnitude below the model scales. In part I of this series of publications \cite{Koenigstein:2021syz}, we discussed that large/low UV/IR cutoffs are needed to ensure cutoff independence of the IR effective action, which is also known as RG consistency \cite{Braun:2018svj}. Interestingly, we find that the almost total absence of a plateau in the entropy in the UV of our first case implies that we almost violated RG consistency.\footnote{The absence of the zero-entropy plateau can also be seen by closer inspection of Fig.~16 of Ref.~\cite{Koenigstein:2021syz}, where $\Lambda = 10^6$ is barely on the plateau of RG consistent UV scales.} For all other test cases this is avoided by choosing larger UV cutoffs $\Lambda$, see below.
	
	Before we continue with our next test case, we again note that the absolute value of $\mathcal{C} [\partial_x u ( t, x ) ]$ in the IR in Fig.~\ref{fig:sc_i_on=1_n=800_xmax=10_lambda=1.0e6_tir=60_entropy_flow} has no quantitative meaning, due to the ill-conditioned behavior when applied to the discontinuous initial condition \eqref{eq:testing_scenario_non-analytic_quadaratic_asymptote} of the numerical derivative \eqref{eq:centeredDifferences}. 
	However, this does not spoil our qualitative arguments at all. 
		
\subsection{Test case II: \texorpdfstring{$\phi^4$}{phi4}-theory} 
\label{subsec:scenario_II}

	The second test case is the zero-dimensional analogue of higher-dimensional $\phi^4$-models. We consider two UV initial conditions, differing in the sign of the mass-like $\phi^2$-contribution,
		\begin{align}
			U ( \sigma ) = \mp \tfrac{1}{2} \, \sigma^2 + \tfrac{1}{4!} \, \sigma^4 \, .	\label{eq:testing_scenario_phi4}
		\end{align}
	Hence, depending on the sign, we either start the RG flow with a broken $\mathbb{Z}_2$-symmetry in the ground state or with a $\mathbb{Z}_2$-symmetric ground state. For a visualization of the initial condition with negative mass term, see Fig.~18 of Ref.~\cite{Koenigstein:2021syz}\footnote{RG flows for both choices of sign within a low-order FRG Taylor expansion are also provided in, \textit{e.g.}, Refs.~\cite{Pawlowski:talk,Keitel:2011pn,Moroz:2011thesis} for the zero-dimensional $O(N)$ model.}. This initial condition is chosen because of its relevance in higher dimensions, \textit{e.g.}, for  
	studies of spontaneous symmetry breaking and symmetry restoration (ranging from applications in statistical mechanics and condensed-matter theory to high-energy physics). Additionally, in contrast to our first test case \eqref{eq:testing_scenario_non-analytic_quadaratic_asymptote}, due to the analyticity of Eq.~\eqref{eq:testing_scenario_phi4}, a generic expansion of the potential in polynomials at any $\sigma$ is possible in the UV at $t = 0$. This property can be used to study the convergence of a common FRG truncation scheme, the Taylor expansion of the effective action. In Sec.~V of Ref.~\cite{Koenigstein:2021syz}, we find that only for positive mass-like terms, where the physical point does not move during the RG flow, the FRG Taylor expansion about the IR minimum $\sigma = 0$ exhibits ``apparent'' convergence by increasing the expansion order. For negative mass terms (also using a fixed expansion point at the IR minimum $\sigma = 0$), we do not find convergence while increasing the expansion order. In Sec.~V of Ref.~\cite{Koenigstein:2021syz}, we argue that during the RG flow, while the physical point moves from $\sigma = \pm \sqrt{6}$ to $\sigma = 0$, presumably an excessively large or even infinitely many new couplings are generated in $U ( t, \sigma )$. This renders FRG Taylor expansion at a finite order a potentially problematic approximation scheme in such a scenario.
	
\begin{figure}
			\centering
			\includegraphics{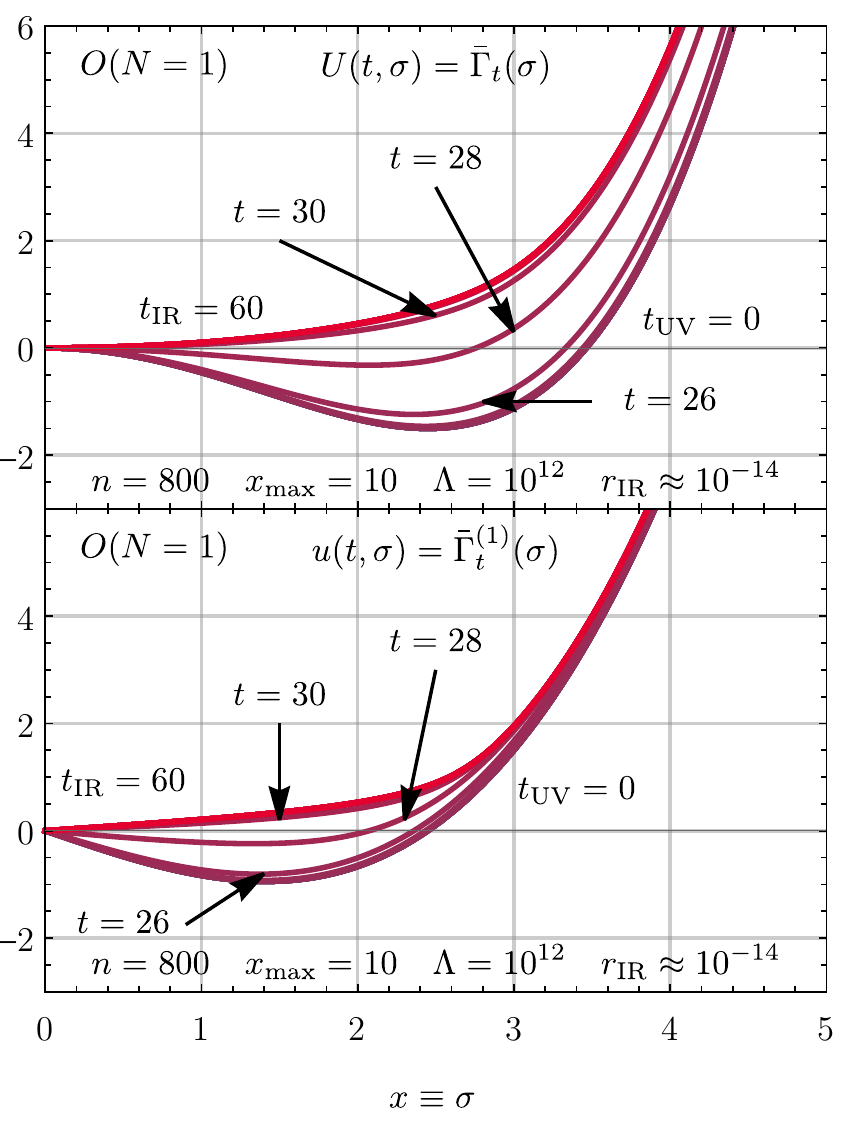}
			\caption{\label{fig:sc_ii_n_on=1_n=800_xmax=10_lambda=1.0e12_tir=60_rg_flow}%
				RG flow of the effective potential $U ( t , \sigma )$ (upper panel) and its derivative $u ( t , \sigma ) = \partial_\sigma U ( t , \sigma )$ (lower panel) for the zero-dimensional $O ( N = 1 )$-model with initial condition Eq.~\eqref{eq:testing_scenario_phi4} (with negative mass term) evaluated at $t = 0, \, 2, \, 4, \, \ldots, \, 60$ (integer values for $t$ were only chosen for convenience and readability). The {blue}/{magenta} curve corresponds to the UV while the {red} curve to the IR. We used the exponential regulator Eq.~\eqref{eq:exponential_regulator} with UV cutoff $\Lambda = 10^{12}$. For convenience only, the plot does not show the region $x = 5$ to $x = 10$ because the tiny differences between $u ( t , \sigma )$ and $u ( t_\mathrm{UV} , \sigma )$ are not visible in this region and vanish for large $x = \sigma$ anyhow.
			}
		\end{figure}

\begin{figure}
			\centering
			\includegraphics{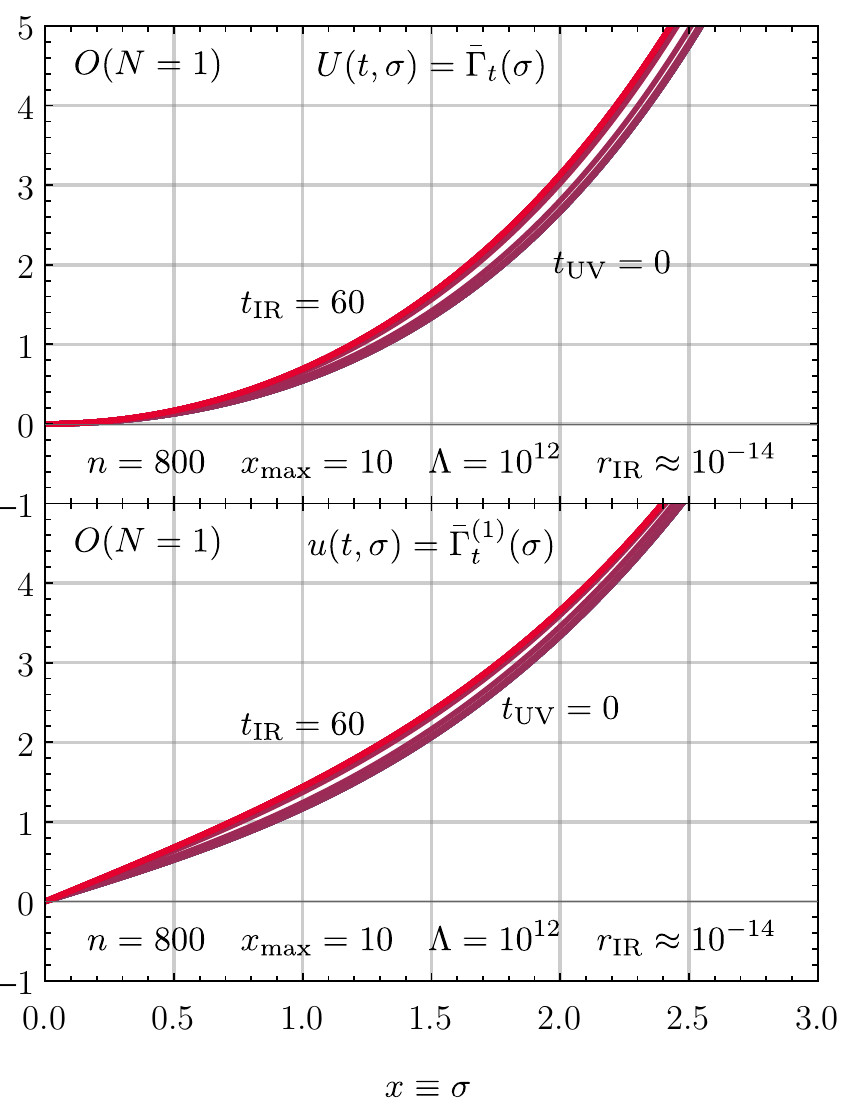}
			\caption{\label{fig:sc_ii_p_on=1_n=800_xmax=10_lambda=1.0e12_tir=60_rg_flow}%
				RG flow of the effective potential $U ( t , \sigma )$ (upper panel) and its derivative $u ( t , \sigma ) = \partial_\sigma U ( t , \sigma )$ (lower panel) for the zero-dimensional $O ( N = 1 )$-model with initial condition Eq.~\eqref{eq:testing_scenario_phi4} (with positive mass term) evaluated at $t = 0, \, 2, \, 4, \, \ldots, \, 60$ (integer values for $t$ were only chosen for convenience and readability). The {blue}/{magenta} curve corresponds to the UV while the {red} curve to the IR. We used the exponential regulator Eq.~\eqref{eq:exponential_regulator} with UV cutoff $\Lambda = 10^{12}$. For convenience only, the plot does not show the region $x = 3$ to $x = 10$ because otherwise the differences between the UV  and the IR curves would not be visible at all for the entire domain.
			}
		\end{figure}

	In this subsection, we reinforce our findings about the non-convergence of expansions of the potential during the RG flow by studying the (numerical) entropy production during the RG flows. The RG flows of $u ( t, x )$ for both initial conditions are depicted in Fig.~\ref{fig:sc_ii_n_on=1_n=800_xmax=10_lambda=1.0e12_tir=60_rg_flow} (for negative mass term) and in Fig.~\ref{fig:sc_ii_p_on=1_n=800_xmax=10_lambda=1.0e12_tir=60_rg_flow} (for positive mass term).\footnote{Note that the plot ranges for the ``positive mass''-case are different from all other plots of RG flows of $u ( t, x )$ in this section. Otherwise, the tiny changes during the RG flow would not be visible at all.}
	
	Both RG flows are by visual inspection not really spectacular: For the ``negative mass''-case, we find that, according to the Coleman-Mermin-Wagner-Hohenberg theorem \cite{Mermin:1966,Hohenberg:1967,Coleman:1973ci}, the diffusion via the $\sigma$-mode restores the $\mathbb{Z}_2$-symmetry and drives the potential convex during the RG flow before the system equilibrates in the IR. For the RG flow of the ``positive mass''-case we only find minimal changes in the shape of $u ( t, x )$ also originating from the non-linear diffusion during the RG flow. Hence, the equilibrated solution in the IR is relatively close to the UV initial potential.
		\begin{figure}
			\centering
			\includegraphics{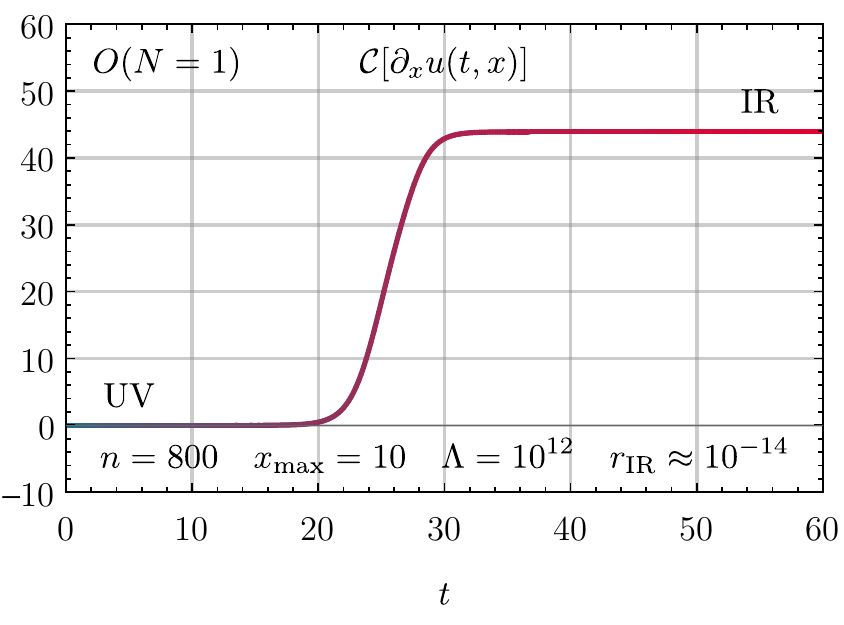}
			\caption{\label{fig:sc_ii_n_on=1_n=800_xmax=10_lambda=1.0e12_tir=60_entropy_flow}%
				The plot shows the monotonic growth of the (numerical) entropy/the $\mathcal{C}$-function $\mathcal{C} [ \partial_x u ( t, x ) ]$ during the RG flow of the test case \eqref{eq:testing_scenario_phi4} (with negative mass term) and corresponds to Fig.~\ref{fig:sc_ii_n_on=1_n=800_xmax=10_lambda=1.0e12_tir=60_rg_flow}.
			}
		\end{figure}
		\begin{figure}
			\centering
			\includegraphics{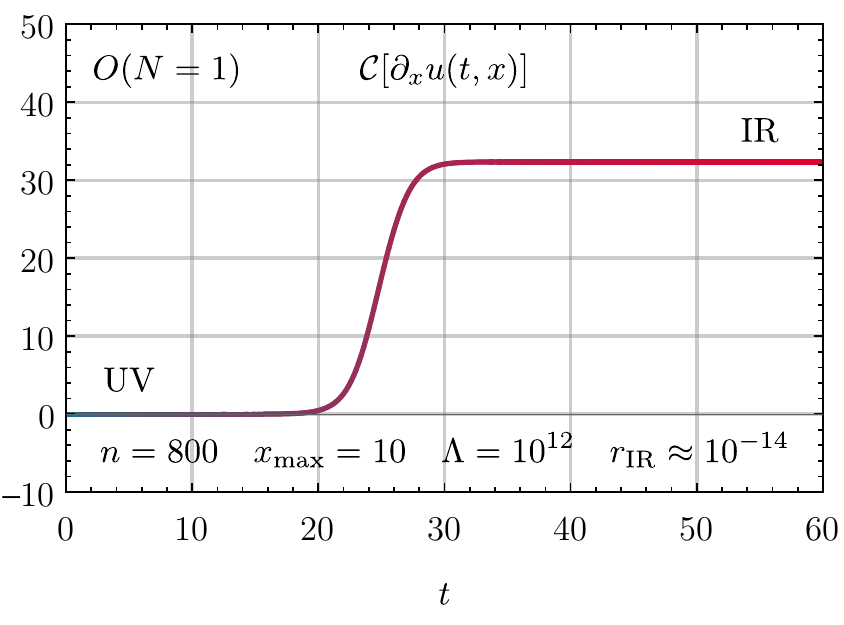}
			\caption{\label{fig:sc_ii_p_on=1_n=800_xmax=10_lambda=1.0e12_tir=60_entropy_flow}%
				The plot shows the monotonic growth of the (numerical) entropy/the $\mathcal{C}$-function $\mathcal{C} [ \partial_x u ( t, x ) ]$ during the RG flow of the test case \eqref{eq:testing_scenario_phi4} (with positive mass term) and corresponds to Fig.~\ref{fig:sc_ii_p_on=1_n=800_xmax=10_lambda=1.0e12_tir=60_rg_flow}.
			}
		\end{figure}
	
	The plots of the corresponding entropies in Fig.~\ref{fig:sc_ii_n_on=1_n=800_xmax=10_lambda=1.0e12_tir=60_entropy_flow} (for negative mass term) and Fig.~\ref{fig:sc_ii_p_on=1_n=800_xmax=10_lambda=1.0e12_tir=60_entropy_flow} (for positive mass term) are more instructive. In both cases we find a clear monotonic rise of the (numerical) entropy exactly in the RG time period, in which most of the dynamics takes place. Furthermore, we clearly find plateaus in the UV and the IR, which correspond to the trivial UV regime and the non-trivial interacting IR regime. This plateau-like behavior signals RG consistency. In comparison with our first test case \eqref{eq:testing_scenario_non-analytic_quadaratic_asymptote}, where we used exactly the same discretization points (volume cells), the monotonic growth of entropy is less drastic and significantly smaller. This is expected because the jumps in $u ( t = 0, x )$ at $x = 2$ and $x = 3$ in the first test case \eqref{eq:testing_scenario_non-analytic_quadaratic_asymptote} lead to greater changes in the discrete total variation [the arc length in $x$ of $u ( t, x )$] than the rather small changes of the profiles of $u ( t, x )$ for the $\phi^4$-models, compare Sub.Sec.~\ref{subsec:discrete_c_and_tvd}. Also from a fluid dynamic perspective, this is intuitively understandable because the smoothening of huge gradients (rarefaction waves) is a substantial source of entropy and obviously an irreversible process, whereas only a small transport of a fluid is not a source of excessive but rather small entropy production, even though it is diffusion driven. Still, also for both $\phi^4$-cases the entropy increases during the RG flow, which first signals an increasing number of coupling constants generated during the RG flow, and second also renders the RG flows irreversible.
	
	The second observation has severe consequences: Any RG flow in a FRG Taylor expansion employs a finite set of coupled ODEs for the couplings (vertices). Since the system is finite, it seems to be theoretically possible integrate in either RG time direction. In higher dimensions, one can formally integrate to larger or lower (energy) scales (associated with resolutions in position space), compare with, \textit{e.g.}, the perturbative $\beta$-functions of QCD, QED \textit{etc.} \cite{Politzer:1973fx,Gross:1973id,Gross:1973ju,Gross:1974cs}. However, this is in principle not compatible with the irreversibility of RG flows as shown in our present work (as, \textit{e.g.}, signaled by the rise of entropy)  and may only be reliable within small subspaces of the theory space associated with a given theory. In fact, the computation of fundamental couplings at small scales (high energies) from effective couplings at large scales (low energies) is in general not possible, \textit{cf.}\ Ref.~\cite{Wilson:1979qg}. We conclude that the increase of entropy, which we also observe during the RG flow of our analytic initial conditions \eqref{eq:testing_scenario_phi4} reveals potential limitations of Taylor expansion of effective actions because most likely an extremely large number (or even infinite number) of couplings is generated in the RG flow and would be required to correctly describe the RG flow. 
	\footnote{%
		At this point, one might be tempted to apply our definition of the normalized (numerical) entropy directly to some $\partial_x u ( t, x )$ that is reconstructed from the flow of the coefficients of a Taylor expansion of the potential to study the validity of the expansion. However, this is not possible, because the FRG Taylor expansion in general provides only an adequate local description of the potential, while our (numerical) entropy or the TV requires knowledge about the global shape of the potential or its derivatives, respectively.
	}%
	\footnote{%
		This reasoning might also resolve some issues, which are discussed in Ref.~\cite{Curtright:2011qg}. In Ref.~\cite{Curtright:2011qg}, it is argued that the ``$\mathcal{C}$-theorem folklore'' about the existence of a monotonically rising $\mathcal{C}$-function prevents the RG flow from entering limiting cycles (or even chaotic behavior) is wrong. However, their arguments are entirely based on examples of $\beta$-functions with a finite number of couplings. As explicitly shown in Ref.~\cite{Curtright:2011qg}, such systems can indeed show limiting cycles and still have monotonic flows \textit{etc.} as shown by the authors. Similar to what is explained at several occasions in our work, such systems do however not show irreversibility in the sense of a true diffusive/dissipative process or via the interaction/generation of discontinuities in field space -- ``the theory-space of couplings'' \cite{Grossi:2019urj,Steil:2021cbu}. The only irreversible character of these systems might indeed be that they can enter limiting cycles, show chaotic behavior or enter fixed points on a finite set of couplings.
	}
	
	In practical computations with the zero-dimensional $O(N)$ model a complete inversion of the RG flow -- integrating up from the IR to the UV -- is possible only under certain conditions. In Sub.Sub.Sec.~V~B~2 of part I of this series of publications \cite{Koenigstein:2021syz} we performed tests with the $\phi^4$ model discussed in this subsection and found that a complete and accurate practical inversion is only possible for the $\phi^4$ model with positive mass term when considering a small set of running couplings. Numerical instabilities related to massive oscillations in the higher-order couplings prevent a numerical inversion of the RG flow for larger systems of couplings. In the $\phi^4$ model with negative mass term a numerical reconstruction of the non-convex UV potential by integrating up from the convex IR potential seems to be practically impossible with the employed Taylor expansions. While an inversion of the RG flow seems theoretically possible on first sight when considering the finite ODE systems of the FRG Taylor (vertex) expansion the practical/numerical realization is not obvious.
	
	When considering an expansion in vertices, it might be possible that higher-order couplings/vertices are strongly suppressed (especially when considering higher dimensional QFTs), 
	such that an expansion of the ERG equation \eqref{eq:exact_renormalization_group_equation} in vertices is applicable and meaningful in practice, see, \textit{e.g.}, Refs.~\cite{Eser:2018jqo,Eser:2019pvd,Divotgey:2019xea,Cichutek:2020bli}. This should go hand in hand with only a small growth of an entropy for the exact RG flow. Exactly this seems to be the case for our ``positive mass'' case \eqref{eq:testing_scenario_phi4}, which shows almost no dynamics at all and yields the smallest increase in entropy of all our test cases. A reason, why here a rather small number of couplings might be sufficient to describe the entire RG flow is that the potential is convex during the entire flow and has a single unique non-moving minimum. Hence, the UV regime of this model and the IR regime do not differ much and, as long as the quartic coupling is extremely small, also perturbation theory \cite{Strocchi:2013awa} leads to results which are consistent with the exact values for the lowest $1$PI-$n$-point-correlation (vertex) functions \cite{Keitel:2011pn}.

\subsection{Test case III: \texorpdfstring{$\phi^6$}{phi6}-potential} 
\label{subsec:scenario_III}

	The third test case describes a potential that is analytic with a $\mathbb{Z}_2$-symmetric ground state in the UV. However, the potential exhibits two non-trivial local minima and behaves asymptotically $\propto \phi^6$,
		\begin{align}
			U ( \sigma ) = \tfrac{1}{2} \, \sigma^2 - \tfrac{1}{20} \, \sigma^4 + \tfrac{1}{6!} \, \sigma^6 \, ,	\label{eq:testing_scenario_phi6}
		\end{align}
	such that it is not convex in the UV. A plot of the UV potential can be found in Fig.~27 of Ref.~\cite{Koenigstein:2021syz}. This initial condition of the RG flow is chosen in Ref.~\cite{Koenigstein:2021syz} to test whether the poor convergence of the FRG Taylor expansion of the $\phi^4$-potential with negative mass term \eqref{eq:testing_scenario_phi4} is merely an artifact of the moving scale-dependent global minimum in the RG flow and to figure out whether the FRG Taylor expansion should have actually be performed at a moving expansion point,  \textit{i.e.}, about the moving global minimum instead of expanding around the IR minimum $\sigma = 0$ during the entire flow. However, although the global minimum at $\sigma = 0$ is not moving at all in the RG flow of the $\phi^6$-case \eqref{eq:testing_scenario_phi6}, the Taylor expansion completely fails here,\footnote{We thank J.~Eser for discussions on this issue and a cross check which reproduced our findings for this test case, using his FRG code for Taylor-expanded effective actions~\cite{Divotgey:2019xea,Cichutek:2020bli,Eser:2018jqo,Eser:2019pvd}.} see Ref.~\cite{Koenigstein:2021syz}. In Ref.~\cite{Koenigstein:2021syz}, we conclude that there has to be a time interval during the RG flow, where $u ( t, x )$ exhibits a highly non-local dynamics. The latter cannot be captured by a local expansion with a finite number of couplings about a single point, even though the expansion point is unique and does not move.
		\begin{figure}
			\centering
			\includegraphics{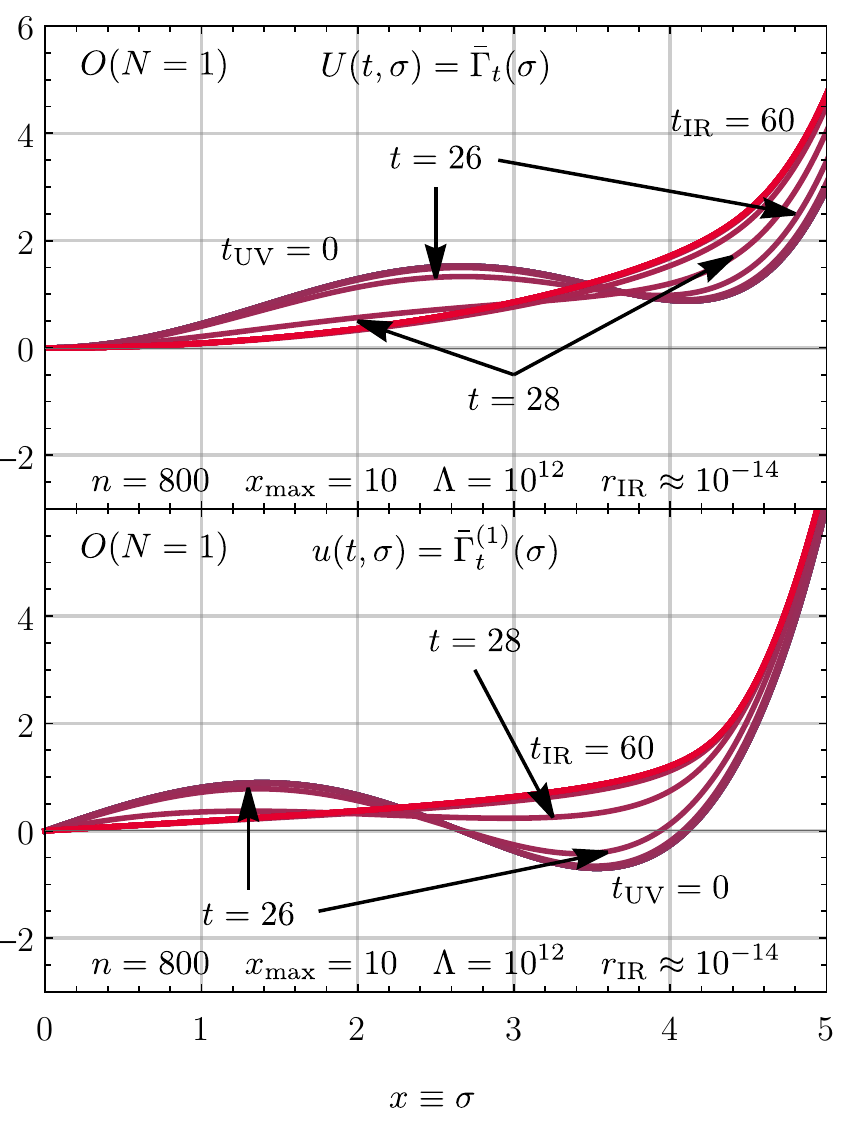}
			\caption{\label{fig:sc_iii_on=1_n=800_xmax=10_lambda=1.0e12_tir=60_rg_flow}%
				RG flow of the effective potential $U ( t , \sigma )$ (upper panel) and its derivative $u ( t , \sigma ) = \partial_\sigma U ( t , \sigma )$ (lower panel) for the zero-dimensional $O ( N = 1 )$-model with initial condition Eq.~\eqref{eq:testing_scenario_phi6} evaluated at $t = 0, \, 2, \, 4, \, \ldots, \, 60$ (integer values for $t$ were only chosen for convenience and readability). The {blue} curve corresponds to the UV while the {red} curve to the IR. We used the exponential regulator Eq.~\eqref{eq:exponential_regulator} with UV cutoff $\Lambda = 10^6$. For convenience only, the plot does not show the region $x = 5$ to $x = 10$ because the tiny differences between $u ( t , \sigma )$ and $u ( t_\mathrm{UV} , \sigma )$ are not visible in this region and vanish for large $x = \sigma$ anyhow.
			}
		\end{figure}
	
	Actually, this can be seen directly from the RG flow of $u ( t, x )$ in Fig.~\ref{fig:sc_iii_on=1_n=800_xmax=10_lambda=1.0e12_tir=60_rg_flow} at approximately $t \approx 27$, which is also the time when the RG flow of the Taylor-/vertex expansion collapses due to strongly oscillating and ultimately diverging couplings at $t \approx 27$. At this RG time, the local minimum (the second non-trivial zero-crossing) vaporizes via the diffusion and merges with the global minimum at $x = 0$. Already from the curves in Fig.~\ref{fig:sc_iii_on=1_n=800_xmax=10_lambda=1.0e12_tir=60_rg_flow} one can observe that the $u ( t, x )$ is hardly describable over the entire RG flow with only a finite set of couplings. The breakdown of any expansion can also be directly related to Wilbraham-Gibbs-oscillations \cite{Wilbraham:1848,Gibbs:1898,Gibbs:1899,boyd2001chebyshev} in the flat region of $u ( t, x )$. This was already (indirectly) described before in the context of FRG studies~\cite{Pangon:2010uf} and represents another direct interplay between characteristic properties of the RG and the numeric treatment of PDEs.
		\begin{figure}
			\centering
			\includegraphics{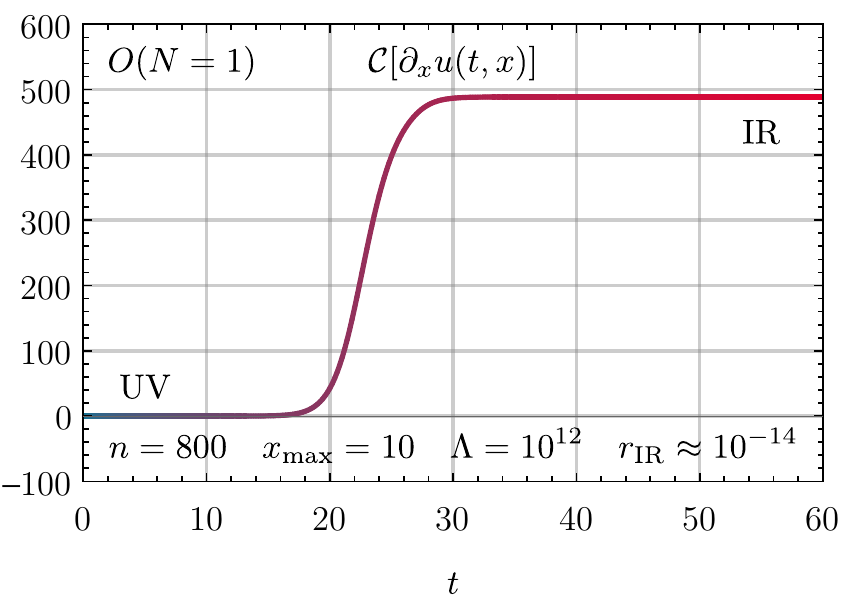}
			\caption{\label{fig:sc_iii_on=1_n=800_xmax=10_lambda=1.0e12_tir=60_entropy_flow}%
				The plot shows the monotonic growth of the (numerical) entropy/the $\mathcal{C}$-function $\mathcal{C} [ \partial_x u ( t, x ) ]$ during the RG flow of the test case \eqref{eq:testing_scenario_phi6} and corresponds to Fig.~\ref{fig:sc_iii_on=1_n=800_xmax=10_lambda=1.0e12_tir=60_rg_flow}
			}
		\end{figure}
	
	Interestingly, also the (numerical) entropy function signals exactly the discussed non-local behavior at $t \approx 27$. Exactly at that point in time when the local minimum merges with the global minimum, we observe the strongest increase of entropy, see Fig.~\ref{fig:sc_iii_on=1_n=800_xmax=10_lambda=1.0e12_tir=60_entropy_flow}. We also find that by absolute measures, the entropy production for the $\phi^6$-initial potential~\eqref{eq:testing_scenario_phi6} is greater than the entropy production observed for both quartic initial conditions~\eqref{eq:testing_scenario_phi4}. Nevertheless, the entropy production for the non-analytic initial condition~\eqref{eq:testing_scenario_non-analytic_quadaratic_asymptote} is still greater than the one in the $\phi^6$-case. This can be easily understood from the relation of the numerical entropy to the total variation, \textit{i.e.}, the arc length of $u ( t, x )$ which even formally diverges for Eq.~\eqref{eq:testing_scenario_non-analytic_quadaratic_asymptote} in the UV.\footnote{Absolute values of the numerical entropy as well as their comparison should be considered with some care as explained above.}
	
	We conclude from this section that the (numerical) entropy might be a tool to detect if the RG flow ``moves" far from the perturbative region, while going from $t = 0$ to $t \rightarrow \infty$. In other words, it is a tool to discuss whether the RG flow is governed by strong (non-perturbative) dynamics and cannot be captured within any kind of local or perturbative expansion. This is analogous to a thermal system or fluid evolving through an out-of-equilibrium state, before finally equilibrating or showing steady flow behavior, in contrast to a temperature distribution or fluid that is already close to its equilibrium state.

\subsection{Test case IV: the \texorpdfstring{$\sigma = 0$}{sigma = 0} boundary}
\label{subsec:scenario_IV}

	Our final test case has originally been used in part I of this series of publications \cite{Koenigstein:2021syz} to test the correct implementation of (spatial) boundary conditions for PDEs of the form~\eqref{eq:flow_equation_conservative}:
		\begin{align}
			U ( \sigma ) =%
			\begin{cases}
				- ( \sigma^2 )^{\tfrac{1}{3}} \, ,		&	\text{if} \quad |\sigma| \leq \sqrt{8} \, ,	\vphantom{\bigg(\bigg)}
				\\
				\tfrac{1}{2} \, \sigma^{\, 2}-6 \, ,	&	\text{if} \quad |\sigma| > \sqrt{8} \, .	\vphantom{\bigg(\bigg)}
			\end{cases}	\label{eq:testing_scenario_4}
		\end{align}
	The UV initial potential $U ( \sigma )$ now exhibits a non-analyticity -- a cusp -- at ${\sigma = 0}$ which leads to a pole in the conserved quantity ${u ( t = 0, \sigma ) = \partial_\sigma U ( \sigma )}$.\footnote{Potentials with cusps in field space can be found in the context of, \textit{e.g.}, theories in 2+1 spacetime dimensions, such as the Gross-Neveu model~\cite{Braun:2010tt}.} Additionally, to put the tests of our numerical approach to the extremes, we incorporated two non-trivial minima at $\sigma = \pm \sqrt{8}$, which are on top of that also non-analytic points, causing again discontinuities in $u ( t = 0, \sigma )$. A visualization of Eq.~\eqref{eq:testing_scenario_4} is shown in Fig.~30 of Ref.~\cite{Koenigstein:2021syz}.

\begin{figure}
			\centering
			\includegraphics{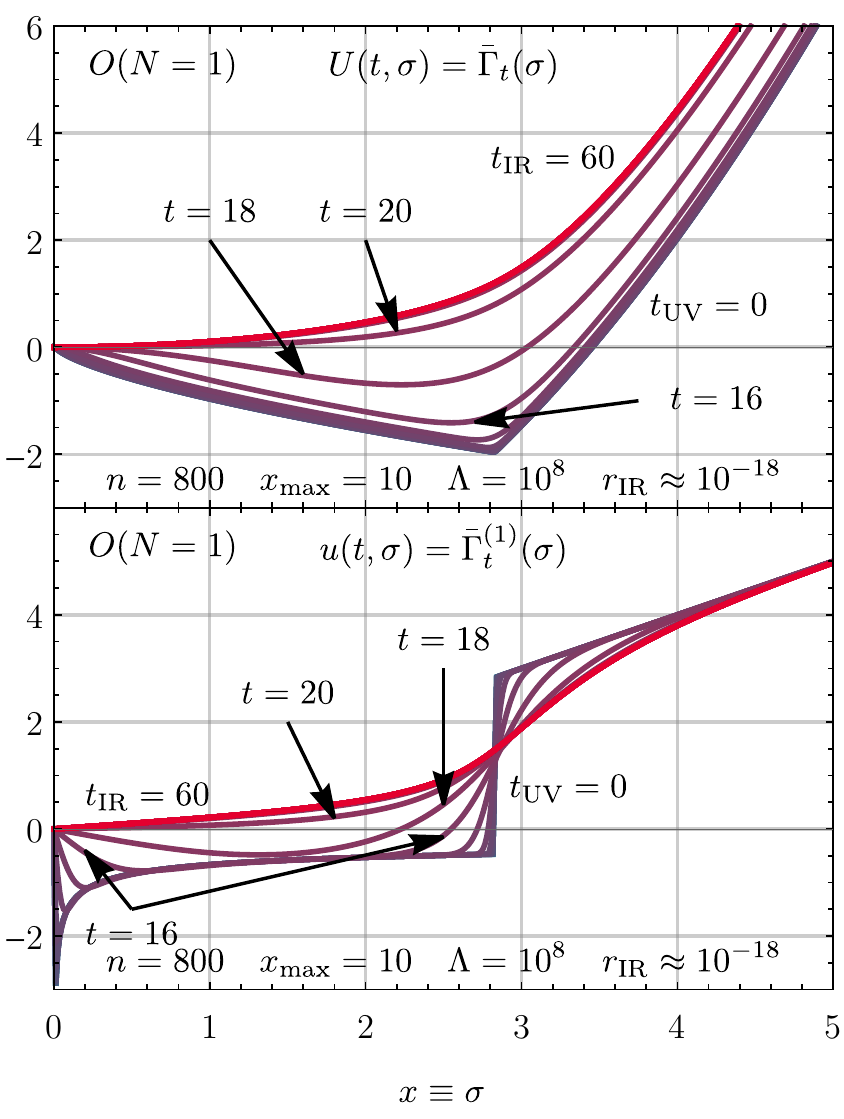}
			\caption{\label{fig:sc_iv_on=1_n=800_xmax=10_lambda=1.0e8_tir=60_rg_flow}%
				The plot shows the RG flow of the effective potential $U ( t , \sigma )$ (upper panel) and its derivative $u ( t , \sigma ) = \partial_\sigma U ( t , \sigma )$ (lower panel) for the zero-dimensional $O ( N = 1 )$-model with initial condition Eq.~\eqref{eq:testing_scenario_4} evaluated at $t = 0, \, 2, \, 4, \, \ldots, \, 60$ (integer values for $t$ were only chosen for convenience and readability). The {blue} curve corresponds to the UV while the {red} curve to the IR. We used the exponential regulator Eq.~\eqref{eq:exponential_regulator} with UV cutoff $\Lambda = 10^6$. For convenience only, the plot does not show the region $x = 5$ to $x = 10$ because the tiny differences between $u ( t , \sigma )$ and $u ( t_\mathrm{UV} , \sigma )$ are not visible in this region and vanish for large $x = \sigma$ anyhow.
			}
		\end{figure}

	Also in the context of this work, the test case~\eqref{eq:testing_scenario_4} turns out to be a highly interesting almost pathological example. The RG flow of $u ( t, x )$ is shown in Fig.~\ref{fig:sc_iv_on=1_n=800_xmax=10_lambda=1.0e8_tir=60_rg_flow}, where one can see that the numerical scheme indeed perfectly copes with the aforementioned somewhat artificial challenges and correctly reproduces symmetry restoration, convexity and smoothness of the potential in the IR regime.
	
	Of specific interest regarding the (numeric) entropy is of course also the pole of $u ( t = 0, x )$ at $x = 0$. Formally, the arc length (the total variation) of $u ( t, x )$, which is directly related to our entropy function, diverges due to the pole at $t = 0$ for all $t > 0$. This divergence is of different nature than the divergence caused by integrating from $x = - \infty$ to $x = + \infty$ in Eq.~\eqref{eq:entropy_func}. Whereas the latter can be cured by normalizing the entropy \textit{w.r.t.}\ the entropy of $u ( t = 0, x)$, the present divergence also occurs on the level of the ``normalized''  entropy function \eqref{eq:Cdef} similar to the other non-analytic jumps in the UV. The reason for the infinite entropy production while going from $t = 0$ to $t > 0$ is exactly that the total variation between $- x_{\mathrm{max}}$ and $+ x_{\mathrm{max}}$ turns finite for $u ( t, x )$ during the flow because the potential becomes convex and smooth. Moreover, symmetry restoration in the ground state sets in for $t \rightarrow \infty$. However, it is still normalized against the infinite total variation of $u ( t = 0, x )$. Interestingly, this problem can be traced back to the initialization of the RG flow equations at $t = 0$ with the classical UV action $\bar{\Gamma}_{t = 0} ( \varphi ) = \mathcal{S} ( \varphi )$, which is actually not totally exact but rather an almost perfect approximation for sufficiently large $\Lambda$, see also our discussion in Ref.~\cite{Koenigstein:2021syz}. In Ref.~\cite{Koenigstein:2021syz}, we argue that the extremely tiny errors stemming from this approximation of the correct initial condition are immediately ``washed out'' after the first RG steps (after an infinitesimal RG time step $\varepsilon > 0$) because of the diffusive character of the ERG equation (as long as $\Lambda$ is chosen sufficiently large). This implies that we can safely ignore the problem of the infinite arc length at $t = 0$, and formally start considering the entropy from $t = \varepsilon$ onward.
		\begin{figure}
			\centering
			\includegraphics{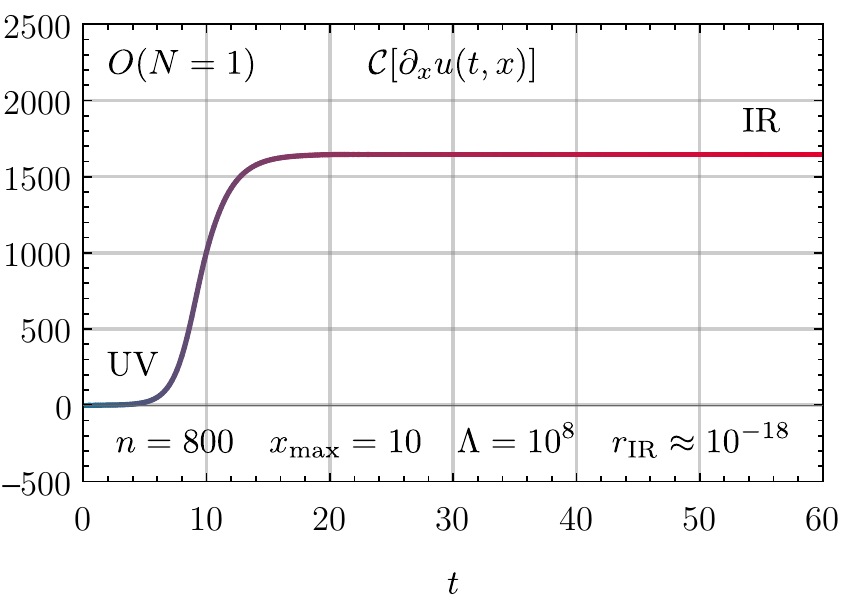}
			\caption{\label{fig:sc_iv_on=1_n=800_xmax=10_lambda=1.0e8_tir=60_entropy_flow}%
				The plot shows the monotonic growth of the (numerical) entropy/the $\mathcal{C}$-function $\mathcal{C} [ \partial_x u ( t, x ) ]$ during the RG flow of the test case \eqref{eq:testing_scenario_4} and corresponds to Fig.~\ref{fig:sc_iv_on=1_n=800_xmax=10_lambda=1.0e8_tir=60_rg_flow}.
			}
		\end{figure}
	
	From a purely practical and numerical perspective, these details may appear somewhat academic anyhow. Via the finite-volume discretization, a smallest resolution $\Delta x$ in field space enters the problem which technically renders the pole at $\sigma = 0$ a huge but already finite jump captured in three volume cells on the level of the cell averages $\bar{u}_i ( t )$ at $t = 0$. Therefore, we can use $u ( t = 0, x )$ as our reference entropy for the normalization of Eq.~\eqref{eq:Cdef} as it is numerically finite right from the beginning of the flow.
	
	An explicit result for the RG flow of our (numerical) entropy is shown in Fig.~\ref{fig:sc_iv_on=1_n=800_xmax=10_lambda=1.0e8_tir=60_entropy_flow}. Irrespective of the subtleties of the preceding discussion, we find a rather large entropy production at exactly those times when the pole vanishes and the jumps at $x = \pm \sqrt{8}$ are smeared out via the diffusion.
	
	Additionally, we find that the total entropy production is much larger for this test case than for the previous ones. Again, this is of course directly related to the huge gradients in the initial condition, which are tremendous sources of entropy via dissipation, directly analogous to the heat equation.\\
	
	In this section, we confronted our theoretical findings with direct numerical computations. We verified the behavior of the function $\mathcal{C} [ \partial_x u ( t, x ) ]$ from Eq.~\eqref{eq:Cdef} by means of its discretized version in Eq.~\eqref{eq:Cdiscrete} as a valid numerical entropy in four test cases. Using the numerical entropy and the ERG equation in the form~\eqref{eq:flow_equation_conservative}, we made several at this point almost intuitive connections between phenomena known in fluid and thermodynamic processes and directly related processes and aspects of RG flows. Most notable, the diffusive character of the flow equation~\eqref{eq:flow_equation_conservative} results directly in irreversible RG flows. This also establishes a connection between steady-state/(thermal) equilibrium solutions and the UV  and IR regime. Moreover, the application of the numerical entropy and total variation appears to be an attractive monitor for RG consistency and the origin of an ``thermodynamic'' time asymmetry.

\section{Irreversibility of the RG flow, entropy and the \texorpdfstring{$\mathcal{C}$}{C}-theorem -- generalizations}
\label{sec:c-theorem_irreversibility_entropy}

	The (re)discoveries within this work unravel the connection between the (numerical) entropy and total variation, employed in applied mathematics, and the irreversibility inherent to RG flows. Furthermore, they might even provide some connections to $\mathcal{C}$-/$\mathcal{A}$-theorems within the framework of truncated RG flow equations. This section is dedicated to a discussion of these aspects of the RG and first generalizations of our findings from our zero-dimensional toy model to higher-dimensional theories.

\subsection{The \texorpdfstring{$\mathcal{C}$}{C}-theorem, fixed points and generalizations to (higher-dimensional) \texorpdfstring{$O(N)$}{O(N)} models}\label{subsec:ctheorem}
	
	The original formulation of the $\mathcal{C}$-theorem \cite{Zamolodchikov:1986gt} states that for a two-dimensional field theory the following properties hold:
		\begin{enumerate}
			\item There exists a positive function 
				\begin{align}
					\mathcal{C} ( \{ g_i \}, t ) \geq 0 \, ,	\label{eq:c-function_1}
				\end{align}
			of all (possibly infinitely many) dimensionless couplings $\{ g_i \}$ of the theory and RG time $t$, with the additional property
				\begin{align}
					\tfrac{\mathrm{d}}{\mathrm{d} t} \, \mathcal{C} ( \{ g_i \}, t ) \geq 0 \, .	\label{eq:c-function_2}
				\end{align}
			(The choice of sign is convention.)
			
			\item The $\mathcal{C}$-function takes a fixed value at (critical) fixed points of the theory, these fixed values can be identified with the central charge $c$ (of the Virasoro algebra)
				\begin{align}
					\mathcal{C} ( \{ g_i^\ast \}, t ) = c \, .	\label{eq:c-function_3}
				\end{align}
			The central charge is different for different fixed points.
		\end{enumerate}
	At first glance, it seems as if our numerical entropy \eqref{eq:Cdef} shares a lot of these properties because all (infinitely many) coupling constants are by definition dimensionless in ${d=0}$ and all included via $u ( t, x )$ in $\mathcal{C} [ \partial_x u ( t, x ) ]$. Furthermore, the function $\mathcal{C} [ \partial_x u ( t, x ) ]$ from Eq.~\eqref{eq:Cdef} monotonically rises during the RG flow and clearly signals irreversibility -- a central aspect of $\mathcal{C}$-/$\mathcal{A}$-theorems \cite{Zamolodchikov:1986gt,Rosten:2010vm,Banks:1987qs,Cardy:1988cwa,Osborn:1989td,Jack:1990eb,Komargodski:2011vj,Curtright:2011qg,Haagensen:1993by,Generowicz:1997he,Forte:1998dx,Codello:2013iqa,Codello:2015ana,Becker:2014pea,Becker:2016zcn}\footnote{Generalizations of Zamolodchikov's $\mathcal{C}$-theorem \cite{Zamolodchikov:1986gt}, especially from two to four dimensions, are often refereed to as $\mathcal{A}$-theorems, referring to Ref.~\cite{Cardy:1988cwa} in which an anomaly coefficient -- hence $\mathcal{A}$-theorem -- in four dimensions is proposed to take the role of the central charge which has given the original $\mathcal{C}$-theorem in two dimensions its name.}. However, there is a crucial difference between our numerical entropy $\mathcal{C} [ \partial_x u ( t, x ) ]$ and the $\mathcal{C}$-function defined via Eqs.~\eqref{eq:c-function_1}-\eqref{eq:c-function_3}, which is more apparent if the discussion is generalized to higher dimensions.

	As already briefly discussed, our previous results should also apply to higher-dimensional $O(1)$ models. This is directly understood by considering the LPA-flow equation of the $d$-dimensional $O(1)$ model using the LPA-optimized regulator \cite{Litim:2000ci,Pawlowski:2017gxj}. The corresponding flow equation for $u ( t, x ) = \partial_x U ( t, x )$ then reads
		\begin{align}
			\partial_t u ( t, x ) = \frac{\mathrm{d}}{\mathrm{d} x} \bigg( - \frac{A_d \, k^{d + 2} ( t )}{k^2 ( t ) + \partial_\sigma u ( t, \sigma )}  \bigg) \, ,	\label{eq:d-dimensional_o(1)}
		\end{align}
	where $A_d = \tfrac{\Omega_d}{d \, (2 \pi)^d}$, and $\Omega_d = \frac{2 \, \pi^{\frac{d}{2}}}{\Gamma ( \frac{d}{2} )}$ is the surface of the $d$-dimensional sphere and the definition~\eqref{eq:rg_time} was used for the RG scale $k(t)$. Apart from numerical and $k$-dependent prefactors, which however do not affect the discussion for the numerical entropy in Sub.Sec.~\ref{subsec:entropy} aside from the caveats already mentioned, the main differences between zero- and higher dimensions are
		\begin{enumerate}
			\item	In the zero-dimensional case, the PDE \eqref{eq:flow_equation_conservative} represents a complete untruncated description of the QFT, while for higher-dimensional $O(1)$ models, the corresponding equation constitutes a truncation of the ERG equation \eqref{eq:exact_renormalization_group_equation}. Still, on the level of the PDE -- thus within the LPA truncation -- the definition of the numerical entropy function \eqref{eq:Cdef} can be used as an entropy and a detector for irreversibility. Additionally, it is not expected that a generic rise of the entropy during the RG flow gets lost, if more sophisticated truncations (like the inclusion of field-dependent wave function renormalizations) are studied because already the ERG equation \eqref{eq:exact_renormalization_group_equation} itself has the form of a non-linear diffusion equation. The numerical entropy \eqref{eq:Cdef} might even help to construct suitable truncation schemes as well as stable numerical methods.
			
			\item	In contrast to the zero-dimensional model, the couplings have in general non-zero energy dimensions. Thus, our numerical entropy \eqref{eq:Cdef} cannot adequately describe the second property of the $\mathcal{C}$-theorem \eqref{eq:c-function_3} -- namely capturing the properties of fixed points, which are defined via the zeroes of the $\beta$-functions of all dimensionless couplings and additionally a constant $\mathcal{C}$-/$\mathcal{A}$-function. To capture the fixed-point structure, one has to replace \eqref{eq:d-dimensional_o(1)} by its rescaled version and derive a corresponding numerical entropy functional.
		\end{enumerate}
			Applying the transformations
				\begin{align}
					y = \, &  A_d^{\frac{1}{2}} \, k^{- \frac{d - 2}{2}} ( t ) \, x \, ,	\vphantom{\bigg(\bigg)}
					\\
					v ( t, y ) = \, & A_d^{\frac{1}{2}} \, k^{-d} ( t ) \, u ( t, x ) \, 	\vphantom{\bigg(\bigg)}
				\end{align}	
			to Eq.~\eqref{eq:d-dimensional_o(1)}, one arrives at the rescaled dimensionless RG flow equation for the derivative of the scale dependent effective potential,
				\begin{align}
					&	\partial_t v ( t, y ) + \tfrac{d}{\mathrm{d} y} \, F [ y, v ( t, y ) ] =	\vphantom{\bigg(\bigg)}	\label{eq:rescaled_flow_equation}
					\\
					= \, & \tfrac{d}{\mathrm{d} y} \, Q [ \partial_y v ( t, y ) ] + S [ v ( t, y ) ] \, ,	\vphantom{\bigg(\bigg)}	\nonumber
				\end{align}
			where
				\begin{align}
					F [ y, v ( t, y ) ] \equiv \, & \tfrac{d - 2}{2} \, y \, v ( t, y ) \, ,	\vphantom{\bigg(\bigg)}	\label{eq:rescaled_advection}
					\\
					Q [ \partial_y v ( t, y ) ] \equiv \, & - \frac{1}{1 + \partial_y v ( t, y )} \, ,	\vphantom{\bigg(\bigg)}	\label{eq:rescaled_diffusion}
					\\
					S [ v ( t, y ) ] \equiv \, & d \, v ( t, y )\, .	\vphantom{\bigg(\bigg)}	\label{eq:rescaled_source}
				\end{align}
			Note that also in its rescaled dimensionless version, the flow equation \eqref{eq:rescaled_flow_equation} for the $d$-dimensional $O(1)$ model has the structure of a non-linear advection-diffusion-source/sink equation, similar to related RG flow equations in  Refs.~\cite{Felder:1987,Zumbach:1993zz,Zumbach:1994kc,Zumbach:1994vg,Hasenfratz:1985dm}, where definition~\eqref{eq:rescaled_advection} corresponds to a position-dependent advection flux,\footnote{For the $O(N)$ model with $N > 1$ the advection flux gains an additional contribution $- ( N - 1)/( 1 + v/y)$.} definition~\eqref{eq:rescaled_diffusion} corresponds to a non-linear diffusion flux, and, finally, definition~\eqref{eq:rescaled_source} corresponds to a source/sink term. We can therefore completely stick to our fluid-dynamic interpretation in terms of conservation laws. Most notably, the diffusive character of the flow equation, thus irreversibility, is also manifest in its rescaled form, independent of the spacetime dimension $d$, see also Refs.~\cite{Felder:1987,Zumbach:1993zz,Zumbach:1994kc,Zumbach:1994vg,Hasenfratz:1985dm} for related RG flow equations.\\
			
			Using Eq.~\eqref{eq:rescaled_flow_equation}, the fixed-point solutions are defined as solutions of the equation with $\partial_t v ( t, x ) = 0$ because all other terms of the flow equation do not explicitly depend on $t$ anymore. A corresponding (numerical) entropy function for Eq.~\eqref{eq:rescaled_flow_equation}, analogously to Eq.~\eqref{eq:Cdef} should therefore signal irreversibility, associated with the change of the number of degrees of freedom for increasing RG time. In particular, it should also assume fixed values at fixed points of the RG flow. We would therefore expect a direct relation between (numerical) entropy in RG flows and Zamolodchikov's formulation~\cite{Zamolodchikov:1986gt} or more recent \cite{Codello:2013iqa,Codello:2015ana} formulations of the $\mathcal{C}-$function. In this respect, we note that it was possible to formulate $\mathcal{C}$-functions for the linearized version of the RG flow equation of the LPA \cite{Zumbach:1993zz,Zumbach:1994kc,Zumbach:1994vg,Rosten:2010vm,Generowicz:1997he}.
			
			Unfortunately, due to the explicitly position-dependent advection flux \eqref{eq:rescaled_flow_equation} as well as the source term \eqref{eq:rescaled_source}, we were not able to formulate an appropriate numerical entropy measure yet.\footnote{%
				The explicit position dependency of the advection flux also prevented us from formulating a numerical entropy for the zero-dimensional $O(N)$ model for finite $N > 1$. The corresponding contribution to the entropy function \eqref{eq:Cdef} allows for $\tfrac{\mathrm{d}}{\mathrm{d} t} \, \mathcal{C} [ \partial_x u ( t, x ) ] < 0$ during RG flow for certain initial conditions and $N$ in the case of $N > 1$. For the zero-dimensional cases \eqref{eq:testing_scenario_phi4} and \eqref{eq:testing_scenario_phi6} discussed in this paper we find $\tfrac{\mathrm{d}}{\mathrm{d} t} \, \mathcal{C} [ \partial_x u ( t, x ) ] < 0$ during the RG evolutions for $N \geq 8$. For the cases \eqref{eq:testing_scenario_non-analytic_quadaratic_asymptote} and \eqref{eq:testing_scenario_4} with their $\sigma^2$ asymptotics for large $\sigma$ the inequality $\tfrac{\mathrm{d}}{\mathrm{d} t} \, \mathcal{C} [ \partial_x u ( t, x ) ] \geq 0$ seems to hold for all $N$ and $t$. We further strongly believe that there are also counterexamples for higher-dimensional $O(N)$ models. Nevertheless, our numerical entropy function \eqref{eq:Cdef} for $N = 1$ might be a good starting point for generalizations to finite $N > 1$, other models and other truncations.%
			} When performing the $y$-derivative of the advection flux~\eqref{eq:rescaled_advection} in Eq.~\eqref{eq:rescaled_flow_equation}, thus considering the advection term in its primitive form, we can distinguish between a purely advective contribution $\propto\partial_y v(t,y)$ and an internal source term $\propto v(t,y)$. Source terms like $S[v(t,y)]$ from Eq.~\eqref{eq:rescaled_source} and from position dependences in non-linear advection or also diffusion fluxes make the construction of numerical entropy functions (like the TV) generally difficult. It is intuitively clear that source and sink terms can change and crucially increase the arc length -- total variation -- of solutions. The construction of numerical entropy functions and related numerical schemes for generic advection-diffusion-source/sink equations is still subject of ongoing research in numerical mathematics \cite{Monthe:2001,Beneito2008,Chen2011May,Bessemoulin:2012}. 
			
			A promising next step might be studies of two-dimensional quantum field theories, where the advection flux \eqref{eq:rescaled_advection} vanishes for the $O(1)$ model also for the rescaled flow equation \eqref{eq:rescaled_flow_equation}. For this case, the system entirely lacks advective contributions and the fields are not rescaled with $k$. Especially the last property might be interesting \textit{w.r.t.}\ the spatial integration contained in the total variation \eqref{eq:TVcontinuous}, which consequently is $k$-independent for both -- the dimensionful and dimensionless -- formulations of the RG flow equation. Furthermore, for $d = 2$ one operates closest to the original version of Zamolodchikov's $\mathcal{C}$-theorem \cite{Zamolodchikov:1986gt}, which should provide some additional guidance to relate the $\mathcal{C}$-theorem to numeric aspects of the PDEs that describe the RG flows.\\
			
			Source terms arising from position-depended advection- (in a formulation in $x$) and diffusion-terms (in a formulation in the invariant $\tfrac{1}{2} \, x^2$) when considering the PDEs in primitive form prevent an obvious generalization of the presented results to zero-dimensional $O(N)$ models with a finite amount $N>1$ of scalars. In App.~E~2~b  of part III of this series of publications \cite{Steil:2021cbu} we discuss numerical entropy and especially total variation/arc length as a candidate for the flow equations of the zero-dimensional $O(N)$ model at finite and notably infinite $N$. Reformulating the RG flow equation in the rescaled invariant ${\tfrac{1}{2N} \, \sigma^2}$ leads to a pure advection equation (with a non-linear but crucially position-independent flux) which are known to be TVNI \cite{HARTEN1983357,Lax1973,Redheffer1974Mar}, thus we identify the difference between initial and current TV as a numerical entropy functional in Eq. (E18) of Ref.~\cite{Steil:2021cbu}.
			
			Formulating a numerical entropy for the flow equation of the zero-dimensional $O(N)$ model at finite $N>1$ might be an important first step toward a formulation in higher dimensions, since it would involve a treatment of source terms $\propto u ( t, x )$. In this sense such a study would be similar to the study in two dimensions proposed in the previous paragraph. Here, a possible starting point might be the observations by Refs.~\cite{Nishigaki:1990sk,DiVecchia:1990ce} that the zero-dimensional analogues of Dyson-Schwinger equations for the $O(N)$ model can be recast into a Virasoro algebra. The Virasoro algebra plays a central role in Zamolodchikov derivation of the $\mathcal{C}$-function for two-dimensional conformal field theories \cite{Zamolodchikov:1986gt}, while the Dyson-Schwinger equations are in direct relation to the RG equations. Also notable in this context is the fact that the Virasoro algebra arising in the study of zero-dimensional $O(N)$ model appears with vanishing central charge -- in this limit strictly speaking as a Witt algebra -- which considering Zamolodchikov definition of a $\mathcal{C}$-function indicates the absence of fixed points in RG flows of the zero-dimensional $O(N)$ model. This will be discussed elsewhere.
	
\subsection{Pointwise monotonicity}

	Interestingly, the discussion of the previous paragraphs shows many similarities with the findings of Refs.~\cite{Becker:2014pea,Becker:2016zcn}. References~\cite{Becker:2014pea,Becker:2016zcn} discuss that for purely bosonic models, the effective average action itself is a ``pointwise monotonic function'', which can be directly seen from the signs of the bosonic contributions in the flow equation \eqref{eq:exact_renormalization_group_equation}\footnote{Related statements about the general monotonicity of solutions of broad classes of PDEs/conservations laws are directly linked to the TVNI property, see \textit{e.g.} Refs.~\cite{HARTEN1983357,LeVeque:1992,Toro2009} and especially Ref.~\cite{Redheffer1974Mar}.}. This implies 
		\begin{align}
			 \partial_t \bar{\Gamma}_k [ \Phi ] \geq 0 \,,\quad \forall \Phi\,. 	\label{eq:pointwise_monotonicity}
		\end{align}
	However, Refs.~\cite{Becker:2014pea,Becker:2016zcn} argue that $\bar{\Gamma}_k [ \Phi ]$ cannot be directly used as a $\mathcal{C}$-function because -- similar to our numerical entropy \eqref{eq:Cdef} -- the above statement is only true for dimensionful field arguments and dimensionful couplings. However, as stated in \cite[p.~3]{Becker:2014pea}, ``the $c$-theorem and its generalizations apply to the RG flow on \textit{theory space}, $\mathcal{T}$, a manifold which is coordinatized by the \textit{dimensionless} couplings. The latter differ from the dimensionful ones by explicit	powers of $k$ fixed by the canonical scaling dimensions. As a consequence, when rewritten	in terms of dimensionless fields and couplings, the property [\eqref{eq:pointwise_monotonicity}] does not precisely translate into a monotonicity statement about the de-dimensionalized theory space analog of [$\bar{\Gamma}_k$], henceforth denoted $\mathcal{A}_k$. Rather, when the derivative $\partial_k$ hits the explicit powers of $k$, additional canonical scaling terms arise which prevent us from concluding simply ``by inspection'' that $\mathcal{A}_k$ is monotonic along RG trajectories.
	
	In fact, we run into the same problems as Refs.~\cite{Becker:2014pea,Becker:2016zcn}. It is the trivial rescaling of dimensionful couplings in terms of the source term \eqref{eq:rescaled_source} that prevents us from immediately writing down an entropy functional or adapted total variation for our flow equation~\eqref{eq:rescaled_flow_equation} (at least for $d = 2$) which could then be directly interpreted as a $\mathcal{C}$-function. In Refs.~\cite{Becker:2014pea,Becker:2016zcn}, it is also the term related to the trivial rescaling which spoils the ``pointwise monotonicity''.

\section{Conclusion and Outlook}\label{sec:conc}

	In this article, we discussed several generic aspects of (F)RG flow equations. We based the discussion of our main findings on the rudimentary example of a zero-dimensional scalar QFT with $\mathbb{Z}_2$-symmetry.
	
	We started off by repeating and deepening the discussion of similarities between RG flow equations and (numerical) fluid dynamics, which was already started in part I of our series of publications \cite{Koenigstein:2021syz}. Based on the formulation of RG flows as advection and diffusion driven dissipative flows in the field space of the corresponding QFT along RG scale (time), we argued that RG flows ``produce" entropy. The RG scale (time) defines a rather natural ``thermodynamic'' arrow of time in this respect. We concluded that this dissipative character of the RG, which causes irreversibility of RG flows, is hard coded in the ERG equation~\eqref{eq:exact_renormalization_group_equation}. This implies that the irreversibility of Kadanoff's block-spin picture is directly encoded in the PDEs (the field dependent $\beta$-functions), which describe the RG flows. Hereby, the IR solutions of RG flows represent equilibrium solutions of fluid dynamic equations. The impossibility of an unambiguous resolution of the microphysics (UV) from the macrophysics (IR) becomes apparent from this standpoint. 
		
	Furthermore, we explicitly demonstrated that the entropy production and the irreversibility during the RG flow from the UV to the IR are not only of abstract manner, but can -- at least for flow equations in certain truncations -- be quantified. Thereby, we directly related the entropy production to the numerical entropy production from the research field of PDEs and numerical fluid dynamics as well as to the total variation non-increasing property. The latter is used to ensure stability of numerical schemes for broad classes of PDEs.
	
	Using our zero-dimensional toy model, we explicitly demonstrated for various test cases taken from our parallel publication \cite{Koenigstein:2021syz} how numerical entropy is produced by diffusion in RG flows and non-analyticities in the UV-initial conditions.
	
	Furthermore, we related certain aspects of the (numeric) entropy production in RG flows to the concept of $\mathcal{C}$-/$\mathcal{A}$-theorems in RG theory since both manifestly encode the irreversible character of RG flows. However, our present study is not yet conclusive in this respect.
	Another interesting aspect related to the introduction of a numeric entropy for RG flows was pointed out by the referee: the present formulation based on the effective potential shares some similarities with the macroscopic description of systems in statistical mechanics. Instead of working with an infinite set of couplings (microstates in statistical mechanics) we switch to a description in terms of an effective potential (a macroscopic formulation in statistical mechanics). The inability (of a macroscopic observer) to track the dynamics of an infinite set of couplings (microstates in statistical mechanics) leads to a macroscopic entropy production/information loss and irreversible processes. An approach to formalize this notion in statistical mechanics was made by Boltzmann \cite{Boltzmann2003} and later Gibbs \cite{Gibbs2010Sep} with the introduction of $H$-theorems, see \textit{e.g.} Chap. VI and XII of the textbook \cite{Tolman1979Nov} for further details. Exploring this connection and possible relations between  $\mathcal{C}$-/$\mathcal{A}$- and $H$-theorems further could be a very interesting prospect for further research.
	
	Although certain aspects of our discussion are still on an abstract level and could not yet be formalized in terms of explicit equations, we believe that our present work provides a fresh view on certain aspects of RG theory, embellished with at at least a few new insights. In particular, our approach may help in the future to construct approximation schemes for RG studies which fully preserve the fundamentally dissipative character of RG flows.

\begin{acknowledgments}
	
	A.K., M.J.S., and J.B. acknowledge the support of the \textit{Deutsche Forschungsgemeinschaft} (DFG, German Research Foundation) through the collaborative research center trans-regio  CRC-TR 211 ``Strong-interaction matter under extreme conditions''-- project number 315477589 -- TRR 211.
	
	A.K. acknowledges the support of the \textit{Friedrich-Naumann-Foundation for Freedom}.
	
	A.K. and M.J.S. acknowledge the support of the \textit{Giersch Foundation} and the \textit{Helmholtz Graduate School for Hadron and Ion Research}.
	
	J.B. acknowledges support by the DFG under Grant No. BR 4005/4-1 and BR 4005/6-1 (Heisenberg program).
	
	E.G. is supported by the U.S. Department of Energy, Office of Science, Office of Nuclear Physics, grants Nos. DE-FG-02-08ER41450.
	
	N.W. is supported by the \textit{Deutsche Forschungsgemeinschaft} (DFG, German Research Foundation) under Germany’s Excellence Strategy EXC 2181/1 - 390900948 (the Heidelberg STRUCTURES Excellence Cluster), the Collaborative Research Centre SFB 1225 (ISOQUANT), the BMBF grant 05P18VHFCA and from the Hessian collaborative research cluster ELEMENTS.\\
	
	We thank the co-authors of the first part of this series of publications \cite{Koenigstein:2021syz}, M.~Buballa and D.~H.~Rischke, for their collaboration, their support, and a lot of exceptionally valuable discussions.
	
	We further thank F.~Divotgey, J.~Eser, F.~Giacosa, F.~Ihssen, L.~Kurth, J.~M.~Pawlowski, K.~Otto, R.~D.~Pisarski, S.~Rechenberger, A.~Sciarra, J.~Stoll, and N.~Zorbach for valuable discussions.\\
	
	We are grateful to the referee of the manuscript for his very insightful comments and questions especially regarding the similarities to statistical mechanics and the $H$-theorems mentioned in the outlook.\\
	
	All numerical results as well as all figures in this work were obtained and designed using \texttt{Mathematica} \cite{Mathematica:12.1} including the following \textit{ResourceFunction}(s) from the \textit{Wolfram Function Repository}: \textit{PlotGrid} \cite{Lang:plotgrid}, \textit{PolygonMarker} \cite{Popkov:polygonmarker}, and \textit{MaTeXInstall} \cite{Horvat:matex}. The ``Feynman'' diagram in Eq.~\eqref{eq:o(1)_wetterich_equation} was generated with \texttt{Axodraw Version 2} \cite{Collins:2016aya}.
\end{acknowledgments}

\bibliography{zero_dim_part_2} 

\end{document}